\documentclass[traditabstract]{aa}  
\usepackage{graphicx}
\begin{document}

   \title{A giant planet in the triple system HD132563
          \thanks{Based on observations made with the Italian Telescopio 
                  Nazionale Galileo (TNG) operated on the island of La Palma 
                  by the Fundacion Galileo Galilei of the INAF 
                  (Istituto Nazionale di Astrofisica) at the Spanish 
                  Observatorio del Roque de los Muchachos of the Instituto 
                  de Astrofisica de Canarias.}}

\titlerunning{A giant planet in the triple system HD132563}

   \author{S. Desidera
          \inst{1},
           E. Carolo
           \inst{1,2},
           R. Gratton
           \inst{1},
           A.F. Martinez Fiorenzano
                        \inst{3},
           M. Endl
           \inst{4},
           D. Mesa
           \inst{1},
           M. Barbieri  
           \inst{5},
          M. Bonavita  
           \inst{6},
           M. Cecconi
           \inst{3},
           R.U. Claudi
          \inst{1},
           R. Cosentino,
          \inst{3,7},
           F. Marzari
          \inst{8},
           \and
           S. Scuderi
           \inst{7}}

   \authorrunning{S. Desidera et al.}

   \offprints{S. Desidera,  \\
              \email{silvano.desidera@oapd.inaf.it} }

   \institute{INAF -- Osservatorio Astronomico di Padova,  
              Vicolo dell' Osservatorio 5, I-35122, Padova, Italy
              \and 
             Dipartimento di Astronomia -- Universit\'a di Padova, Vicolo
             dell'Osservatorio 2, Padova, Italy 
             \and
             Fundaci\'on Galileo Galilei - INAF,
             Rambla Jos\'e Ana Fernández P\'erez, 7
             38712 Bre\~na Baja, TF - Spain
             \and
             McDonald Observatory, The University of Texas at Austin, Austin, 
             TX 78712, USA
             \and
             Observatoire de la Cote d'Azur, Nice, France
             \and
             Department of Astronomy and Astrophysics, University of Toronto, 50 St. George Street Toronto ON Canada
             \and
             INAF -- Osservatorio Astrofisico di Catania, Via S.Sofia 78, Catania, Italy
            \and 
             Dipartimento di Fisica -- Universit\'a di Padova, Via Marzolo 8,
             Padova, Italy }

 \date{Received  / Accepted }

\abstract{
As part of our radial velocity planet-search survey performed with SARG at TNG, we monitored 
the components of HD 132563 for ten years. It is a binary system formed by two rather similar solar type stars
with a projected separation of 4.1 arcsec, which corresponds to 400 AU at the distance of 96 pc.
The two components are moderately metal-poor ([Fe/H]=--0.19), and the age of the system is about 5 Gyr.
We detected RV variations of HD 132563B with period of 1544 days and semi-amplitude of 26 m/s. 
From the star characteristics and line profile measurements, we infer their Keplerian origin.
Therefore HD 132563B turns out to host a planet with a projected mass $m \sin i=1.49~M_{J}$ at 2.6 AU
with a moderately eccentric orbit ($e=0.22$).
The planet around HD 132563B is one of the few that are known in triple stellar systems, as we found that
the primary HD 132563A is itself a spectroscopic binary with a period longer than 15 years and
an eccentricity higher than 0.65. The spectroscopic component was not detected in adaptive-optics images taken with AdOpt@TNG,
since it expected at a projected separation that was smaller than 0.2 arcsec at the time of our observations. 
A small excess in K band difference between the components 
with respect to the difference in V band is compatible with a companion of about $0.55~M{_\odot}$. 
A preliminary statistical analysis of when planets occur in triple systems indicate a similar frequency
of planets around the isolated component in a triple system, components of wide binaries and single stars.
There is no significant iron abundance difference between the components ($\Delta$[Fe/H]=$0.012\pm 0.013$ dex). 
The lack of stars in binary systems and open clusters
showing strong enhancements of iron abundance, which are comparable to the
typical metallicity difference between stars with and without giant planets,
agrees with the idea that accretion of planetary material producing iron abundance anomalies over 0.1 dex is rare. }


   \maketitle
%

\section{Introduction}
\label{s:intro}

Planets in multiple stellar systems play a special role in
determinations of the frequency of planets in the Galaxy,
because more than half of solar type stars are in multiple systems
(Duquennoy \& Mayor \cite{duq}).
Furthermore, planets in binaries allow one to better understand the
processes of planet formation and evolution, thanks to the dynamical
effects induced by the stellar companions on the protoplanetary disks
and on an already formed planetary system (Marzari et al.~\cite{marzari10}).
Finally, since the frequency of (wide) binaries seems related to the density
of the formation environment (Lada \& Lada \cite{lada03}; Goodwin \cite{goodwin10}) 
some insight into the impact of this on the planet formation mechanism
can be derived (Malmberg et al.~\cite{malmberg07}).

In spite of selection criteria against binaries in many planet-search surveys,
several tens of planets are currently known in multiple systems
(Desidera \& Barbieri \cite{db07}, Mugrauer \& Neuhauser \cite{mugrauer09}).
A variety of binary configurations are observed: several of the
planets are in rather wide binaries, for which the effects caused by the companion
on the planetary system are expected to be small. However, there are few planet hosts
with companions at a separation that is small enough (e.g. Hatzes et al.~\cite{hatzes03}; 
Chauvin et al.~\cite{chauvin11}) to make it difficult to
allow planet formation according to our current knowledge (see e.g. Thebault \cite{thebault11})
even if they are on dynamically stable orbits. 
The variety of system configurations is further extended by the discovery
of planets in triple systems and, more recently, of planets in circumbinary
orbits around close eclipsing binary systems (e.g Lee et al.~\cite{lee09}), when using the timing technique.

Planet properties also appear to be affected by the presence of companions, as is the case
for the mass distribution, with massive planets in close orbits found primarily
in binaries (Zucker \& Mazeh \cite{zucker02}; Udry et al.~\cite{udry03}; 
Desidera \& Barbieri \cite{db07}), and for planet eccentricity, which is enhanced
in binary systems, possibly through the Kozai resonance (Tamuz et al.~\cite{tamuz08}).
The circumbinary planets found so far around eclipsing binaries are all rather massive and have long periods.
The possible link of these features with the original system configuration and its
evolution (all the central binaries include a remnant object) needs to be investigated.

The general radial velocity (RV) surveys include several binaries in their samples, but in most cases
only the brightest star in the system is under RV monitoring.
Dedicated surveys of planets in binary systems with well-defined
properties help for better understanding the impact of binarity on planet formation.
Some of these projects are focused on spectroscopic binaries (Eggenberger \cite{egg09};
Konacki et al.~\cite{konacki}), while others study visual binaries using RV (our
survey,  Desidera et al.~\cite{sargbook}; Toyota et al.~\cite{toyota09})
and astrometry (Muterspaugh et al.~\cite{muterspaugh10}).
Our ongoing survey using SARG at TNG is the first planet-search survey specifically devoted to binary systems. It 
targets moderately wide binaries (typical separation 100-500 AU) with similar components (twins).
The sample includes about 50 pairs of main sequence solar-type stars with projected separation larger
than 2 arcsec, distance smaller than 100 pc and magnitude difference $\Delta V < 1$ mag.
The sample selection and the survey itself are described in Desidera et al.~(\cite{sargbook}).

Unlike most of the multiple systems included in the RV surveys, both 
components of the pair are systematically observed in our survey, usually with the same time sampling.
This allows the investigation of possible factors driving planets
in binary systems to occur and of possible effects induced, directly or indirectly, by the presence of planets, such
as altering chemical abundances by accreting metal-rich planetary material (Desidera et al.~\cite{chem2}).

In this paper we present the interesting case of the \object{HD 132563} system, whose components
show both RV variations: those with amplitudes larger than 1 km/s for the primary, implying most likely a 
stellar companion, and those with amplitudes of a few tens of m/s for the secondary, suggesting there is
a planetary companion. The paper is organized as follows.
In Sect.~\ref{s:obs} we present the observations taken at TNG using SARG (high-resolution spectroscopy)
and AdOpt (direct imaging). In Sect.~\ref{s:star} we present the properties of the components.
In Sect.~\ref{s:hd132563b} we present the evidence for a planetary companion around \object{HD132563B}.
In Sect.~\ref{s:hd132563a} we show that \object{HD132563A} is a long-period spectroscopic binary, and we discuss 
the clues to the mass and the orbital parameters of the companion. In Sect.~\ref{s:binorbit} we present
the clues to the orbital parameters of the wide pair and its possible influence on the planet orbiting
HD 132563B. In Sect.~\ref{s:discussion} we discuss the results, with emphasis on planets
in triple systems and on the metallicity difference between components with and without planets. 
In Sect.~\ref{s:conclusion} we summarize our conclusions.

\section{Observations and data reduction}
\label{s:obs}

\subsection{High-resolution spectroscopy with SARG}
\label{s:sarg}

SARG, the high-resolution spectrograph of TNG (Gratton et al.~\cite{sarg}), was used
to gather high-resolution spectra of the components of HD132563.
The spectrograph is equipped with an iodine cell to obtain high-precision
radial velocities. 
The spectra were taken with the yellow grism (spectral coverage between 4620 and 7920~\AA), 
which is ideal for including the full range over which the lines of the iodine cell are present
and which at the same time provides
the highest instrument efficiency. We used The 0.25 arcsec slit, which
yields a 2-pixel spectral resolution of 144000.
All the spectra except one (used as template for the RV determination and for
stellar characterization purposes) were obtained
with the iodine cell inserted into the optical path.

Integration time was fixed at 900 or 1200 s, to keep errors due to the lack of
knowledge of flux mid-time of each exposure smaller than photon noise.
After HD132563B was selected as a high-priority planet candidate, two consecutive
spectra were acquired to make RV errors smaller.
In these case, nightly averages were computed and used in the following analysis.
The templates were obtained with an integration time of 3600 s in optimal seeing conditions
(0.75 arcsec).
Median seeing of star+iodine spectra, measured as FWHM of the spectra across dispersion, is 1.23 arcsec.
While the contamination of the spectra is not of particular concern given
the projected separation of this pair (4.1 arcsec), we followed our standard
observing procedure that foresees a slit oriented perpendicularly to the binary projected separation
to minimize the contamination of the spectra by the other
component.
Overall, 53 and  47 spectra with the iodine cell of HD132563 A and B respectively were
acquired from June 2001 to April 2011.
Data reduction was performed in a standard way using IRAF.
Radial velocities were obtained using the AUSTRAL code (Endl et al.~\cite{austral}),
as described in Desidera et al.~(\cite{hd219542}).

\subsection{Imaging with Adopt@TNG}
\label{s:adopt}

\object{HD 132563} was observed with AdOpt@TNG, the adaptive optics module of TNG
(Cecconi et al.~\cite{adopt}), with the goal of detecting or constraining the mass
of the companion responsible of the RV variations of HD132563A and to better characterize 
the circumstellar environment of both components. 
HD 132563 was observed on 20 Jul 2007  (a series of 22 images) 
and on 10 Aug 2008 using the Br$\gamma$ intermediate-band filter (three series of 27, 23, and 29 images at 
different rotation angles to better remove instrumental artifacts).
Individual integration times were chosen to avoid detector saturation from the
bright components, as we are aiming to look for companions at small angular separation, where
the responsible for the RV variations of HD132563a is expected (see Sect.~\ref{s:montecarlo}). 
Pointing jitter was performed for sky subtraction purposes. 
The primary was used as reference star for adaptive optics.

The instrument has a field of view of about 44 $\times$ 44 arcsec, with a
pixel scale of 0.0437 \arcsec/pixel. Plate scale and absolute detector
orientation were derived by us when observing the systems with
long term trends and planet candidates of our survey 
(see Desidera et al.~\cite{sargbook} for some preliminary results).


Data reduction was performed by first correcting for detector crosstalk
using routines developed for this purpose at 
TNG\footnote{\tt www.tng.iac.es/instruments/nics/files/crt\_nics7.f\,.}
and then performing standard image preprocessing (flat-fielding, bad pixels,
and sky corrections) in the IRAF environment. 

\begin{figure}
   \includegraphics[width=9cm]{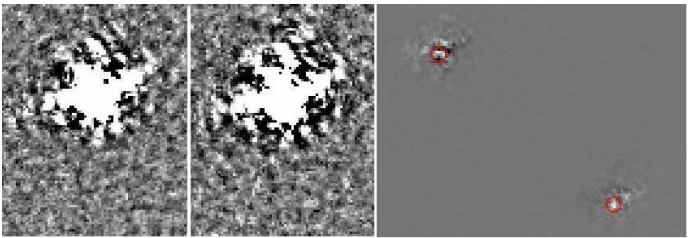}
      \caption{Image of HD132563 system obtained with AdOpt@TNG. Left and central panels: HD 132563 A and B, respectively.
               Right panel: image obtained subtracting the companion PSF
               for each component. The red circles show the 1 $\lambda/D$ limit for each component.}        
         \label{f:image}
   \end{figure}

The two components are within the isoplanatic angle and then characterized
by a very similar PSF (Figure~\ref{f:image}). This allows us to achieve much better detection limits
on difference images, subtracting the PSF of one component from the other.

Images with the same rotation angle were
reduced together to obtain a single image for each different rotation. As a first step we created two separate images
by masking one of the components. On these images we found the position of the center
of the component that is not masked by applying a two-dimensional Gaussian fit procedure. We then subtracted the stellar
profiles of both the components by subtracting the standard deviation in circular annuli around the position 
of  the star. We then
applied a procedure to remove bad and hot pixels that simply substitute the value of a pixel with the median of the eight
surrounding ones if its value is higher than the value of the median plus four times the standard deviation on these pixels.
All the images were then shifted so that the position of the stars is always the same. 
Finally, we applied a high-pass filtering
procedure that substitutes the value of each pixel with the median of a square of $n\times n$ pixels around it, where
however, we excluded the central pixels to avoid subtracting eventual objects. The value of $n$ can be changed according to
the distance from the central star. At the end of the procedure, we made a mean of the the two masked images to get the
final reduced image. We then made a median on the datacubes containing all the images with the same rotation. 
We did discard the
images with a value of the FWHM that was higher than the value of the median plus the standard deviation of the distribution.
When images taken at different rotation angles are available, the resulting images from this procedure were then rotated 
and summed together to enhance the background objects in the FOV. 
To further reduce the speckle noise in the final image we subtracted each other the square regions composed by
$51\times 51$ pixels around the position of the two components. Before the subtraction the two subimages were
normalized, taking their total flux into account. 
Figure~\ref{f:image} shows the image after subtracting of the two components
of the HD132563 system.  Table \ref{t:adopt} lists our measurements of the relative astrometry
between the components.

The magnitude difference in the $Br_{\gamma}$ band was obtained through aperture photometry. Internal errors are within 0.005 mag. 
Additional errors due to transformation on standard system
cannot be determined from our observations. However, we expect the effect on magnitude difference to be small,
thanks to the similar colors of the stars. We conservatively adopt an error of 0.010 mag.

\begin{table}[h]
   \caption[]{Relative astrometry of the components of HD 132563 
   from the observations with AdOpt@TNG. }
     \label{t:adopt}

       \begin{tabular}{lcc}
         \hline
         \noalign{\smallskip}
Epoch   &  $\theta$  & $\rho$   \\
        &  deg &   arcsec       \\
         \noalign{\smallskip}
         \hline
         \noalign{\smallskip}
2007.548  &    276.69$\pm$0.12 &    4.109$\pm$0.015  \\
2008.608  &    276.95$\pm$0.16 &    4.130$\pm$0.015 \\
         \noalign{\smallskip}
         \hline
      \end{tabular}
\end{table}
      
\section{Stellar properties}
\label{s:star}

\object{HD 132563} (HIP 73261, ADS 9461) is a wide pair of main
sequence, solar-type stars.
The physical associations of the components was confirmed by the 
Hipparcos and historical astrometry, the small RV difference between the
components, and the spectral properties of the objects (see below).
The main properties of the components of HD 132563 are summarized in 
Table~\ref{t:star_param}.
An additional component (\object{ADS 9461 C}) at about 1 
arcmin is listed in several binary star catalogs (e.g. CCDM, WDS). 
However it is not physically associated to the system (different proper motion with respect
to HD 132563AB; Roeser et al.~\cite{roeser08}).
We note that the photometry of the individual components likely has larger errors  than
those quoted in Table \ref{t:star_param}, because of blending effects.

\begin{table}[h]
   \caption[]{Stellar properties of the components of HD 132563; where one value is listed at the
               center of the table means the measurement refers to joined A+B components or to the difference
               between the components.}
     \label{t:star_param}

       \begin{tabular}{lccc}
         \hline
         \noalign{\smallskip}
         Parameter   &  HD 132563 A & HD~132563~B  & Ref. \\
         \noalign{\smallskip}
         \hline
         \noalign{\smallskip}

$\alpha$ (2000)          &  14 58 21.519 &  14 58 21.136&  1   \\
$\delta$ (2000)          & +44 02 35.33 &  +44 02 35.87 &  1   \\
$\mu_{\alpha}$ (mas/yr)  & \multicolumn{2}{c}{-59.86 $\pm$ 1.41}  & 2   \\
$\mu_{\delta}$ (mas/yr)  & \multicolumn{2}{c}{-70.12 $\pm$ 1.19}  & 2   \\
RV     (km/s)            &  -6.68$\pm$ 0.20$^{\mathrm{a}}$ & -6.05$\pm$0.20 & 3 \\
RV     (km/s)            &  \multicolumn{2}{c}{-6.5$\pm$1.0$^{\mathrm{b}}$}   & 3 \\
$\pi$  (mas)             & \multicolumn{2}{c}{10.41 $\pm$ 1.45}   & 2   \\
$d$    (pc)              & \multicolumn{2}{c}{96 $\pm$  13}  & 2 \\
 & &  &   \\
$U$   (km/s)             & \multicolumn{2}{c}{8.8$\pm$1.1}  & 3 \\
$V$   (km/s)             & \multicolumn{2}{c}{-39.3$\pm$4.9}  & 3 \\
$W$   (km/s)             & \multicolumn{2}{c}{13.5$\pm$2.8}  & 3 \\
$r_{min}$   (kpc)        & \multicolumn{2}{c}{6.30$\pm$0.31}  & 3 \\
$r_{max}$   (kpc)        & \multicolumn{2}{c}{8.594$\pm$0.024}  & 3 \\
$z_{max}$   (kpc)        & \multicolumn{2}{c}{0.277$\pm$0.039}  & 3 \\
$e$                      & \multicolumn{2}{c}{0.154$\pm$0.023}  & 3 \\
 & &  &   \\
V                        &  8.966$\pm$0.007  &   9.472$\pm$0.010  & 4 \\
$\Delta V$               & \multicolumn{2}{c}{  0.505$\pm$0.018  }  & 4 \\
B-V                      & \multicolumn{2}{c}{  0.560$\pm$0.023  }  &  1 \\
B-V                      &   0.54$\pm$.03    &    0.57$\pm$0.05   & 4 \\
V-I                      &      \multicolumn{2}{c}{  0.63$\pm$0.02  }    & 1 \\
$H_{p}$ scatter          & \multicolumn{2}{c}{0.016} & 1 \\
$J_{2Mass}$              &              &    7.731$\pm$0.054       & 5 \\
$H_{2Mass}$              &              &    7.576$\pm$0.033       & 5 \\
$K_{2Mass}$              &  7.431$\pm$0.21  &    7.494$\pm$0.049       & 5 \\
$\Delta K$               & \multicolumn{2}{c}{  0.467$\pm$0.010  }  & 3 \\
NUV magnitude            &     \multicolumn{2}{c}{13.465$\pm$0.003}  & 6\\
FUV magnitude            &     \multicolumn{2}{c}{19.27$\pm$0.08}  & 6\\
 & & &  \\
$M_{V}$                  &   4.05$\pm$0.30   &   4.56$\pm$0.30       & 3 \\
$T_{eff}$ (K)            &  6168$\pm$100$^{\mathrm{c}}$ &  5985$\pm$100    &  4 \\
$\Delta T_{eff}(A-B)$ (K)& \multicolumn{2}{c}{   184$\pm$12   } &  4 \\
$\log g$                 &  4.15$^{\mathrm{c}}$    &   4.27      &  4 \\
 & & &    \\
$ v \sin i $ (km/s)      & 2.0$\pm$0.8$^{\mathrm{c}}$   &  1.5$\pm$0.5     &  3 \\
 & & &   \\
${\rm [Fe/H]}$           & -0.18 $\pm$ 0.10$^{\mathrm{c}}$ &  -0.19 $\pm$ 0.10 &  4 \\
$\Delta {\rm [Fe/H]}(A-B)$& \multicolumn{2}{c}{0.012 $\pm$ 0.013 } &  4 \\
EW Li 6707 \AA           &   30.2$^{\mathrm{c}}$  &  42.9        &  3 \\
$\log N_{Li}$            &   2.7$^{\mathrm{c}}$   &    2.7      &  3 \\
 & & &   \\
${\rm Mass} (M_{\odot})$ & 1.081$\pm$ 0.010$^{\mathrm{c}}$ & 1.010 $\pm$ 0.010 &  3 \\
Age  (Gyr)               &\multicolumn{2}{c}{$\sim 5$}  &  3  \\

         \noalign{\smallskip}
         \hline
      \end{tabular}

References: 1 Hipparcos (ESA \cite{hipparcos});
            2 van Leeuwen (\cite{newhip}); 
            3 This Paper;
            4 Desidera et al.~(\cite{chem2})
            5 2MASS (Skrutskie et al.~\cite{2mass});
            6 GALEX (Martin et al.~\cite{galex}).

\begin{list}{}{}
\item[$^{\mathrm{a}}$] Instantaneous RV of HD132563Aa 
\item[$^{\mathrm{b}}$] Approximate system RV taking into account all the known components in
                       the system
\item[$^{\mathrm{c}}$] The value refers to HD132563Aa component only.
\end{list}
\end{table}

\subsection{Chemical composition}  
\label{s:chem}

The iron abundance of the components of HD132563 was determined in 
Desidera et al.~(\cite{chem2}). The stars result moderately metal-poor
([Fe/H]=-0.19). There is no significant iron abundance difference
between the components ($\Delta $[Fe/H]$=0.013\pm0.012$).
The implication of this result is
discussed further in Sect.~\ref{s:diffabu}.
The abundance of $\alpha$ elements compared to iron is close to solar, as expected
for thin-disk stars.

Lithium abundance was derived through spectral synthesis of the 6708~\AA~Li doublet, 
adopting the atmospheric parameters from Desidera et al.~(\cite{chem2}).
We found $\log N_{Li}=2.7 \pm 0.1 $ for HD 132563A and 
$\log N_{Li}=2.7 \pm 0.1 $ for HD 132563B.

\subsection{Isochrone fitting and stellar masses}
\label{s:isoc}

\begin{figure}
   \includegraphics[width=9cm]{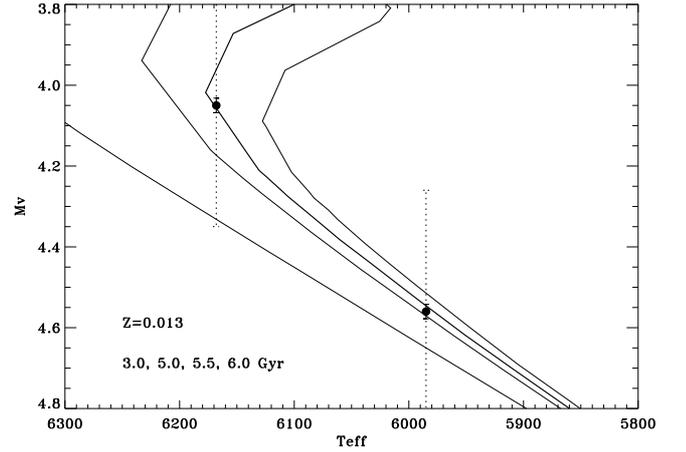}
      \caption{Isochrone fitting of HD132563 system. The Girardi et al.~(\cite{girardi02}) 
               3.0, 5.0, 5.5, and 6.0 Gyr Z=0.013 (the nominal metallicity of the
              system) isochrones are shown, together with the observational data. 
              The larger error bars (dotted lines) refer to the absolute error
               (dominated by error on distance); the smaller error bars (continuous lines) refer to errors on the difference
              between the components. }
         \label{f:cmdv}
   \end{figure}

After we adopted the stellar parameters from the abundance analysis, the stellar models
by Girardi et al.~(\cite{girardi02}; using the web interface param\footnote{\tt http://stev.oapd.inaf.it/param}, 
Da Silva et al.~\cite{param})
yield stellar masses of $1.071\pm0.045$ for \object{HD 132563A} and $0.991\pm0.046~M_{\odot}$ for \object{HD 132563B} 
and ages of $3.6\pm3.0$ and $4.4\pm3.9$ Gyr.
However, these estimates do not consider the much smaller error bars on the difference of magnitude, temperature, and
metallicity between the components. The magnitude difference $\Delta V$, coupled with the effective temperature
difference $\Delta T_{\rm eff}$, indicates that the primary is slightly evolved.  
After adopting the spectroscopic effective temperatures and metallicities, 
the age is tightly constrained at 5.5 Gyr (Fig.~\ref{f:cmdv}).
Uncertainties in temperature scale and absolute abundance implies an error of about 1.5 Gyr on stellar age.
The stellar masses determined from the best-fit isochrone\footnote{Generated using the web tools available at 
\tt http://stev.oapd.inaf.it/cmd}
are 1.081 and 1.010 $M_{\odot}$ for HD 132563 A and B, respectively.

\subsection{Activity and rotation}
\label{s:activity}

Projected rotational velocities were measured by fast Fourier transform (FFT) of absorption profiles, 
which were computed by the cross correlation function (CCF) of the stellar spectra involved in the line bisector study.
About 30-35 unblended lines, mainly Fe in the wavelength range 5230-6050~\AA, were selected when building the CCF.
The CCF profiles (i.e. the observed profiles from the stars) were made symmetric by mirroring 
one of its halves, aiming to reduce the noise of the FFT. 
A new profile was calculated by the convolution of a macroturbulence profile 
(Gaussian) and a rotational one, to compare the FFTs of the symmetric
and the calculated (model) profiles. Macroturbulence was estimated from effective temperature 
using the standard relation by Valenti \& Fischer (\cite{vf05}).
The projected rotation velocity $v \sin i$ corresponds to the case where the first minimum of the FFT 
from the calculated profile coincides with the first minimum of the FFT from the symmetric one, 
as described by Gray (\cite{gray}).
The projected rotation velocities of HD 132563 A and B result 
$v \sin i  = 2.0\pm0.8$ and $1.5\pm0.5$~km/s, respectively. 

 \begin{figure}
   \includegraphics[width=9cm]{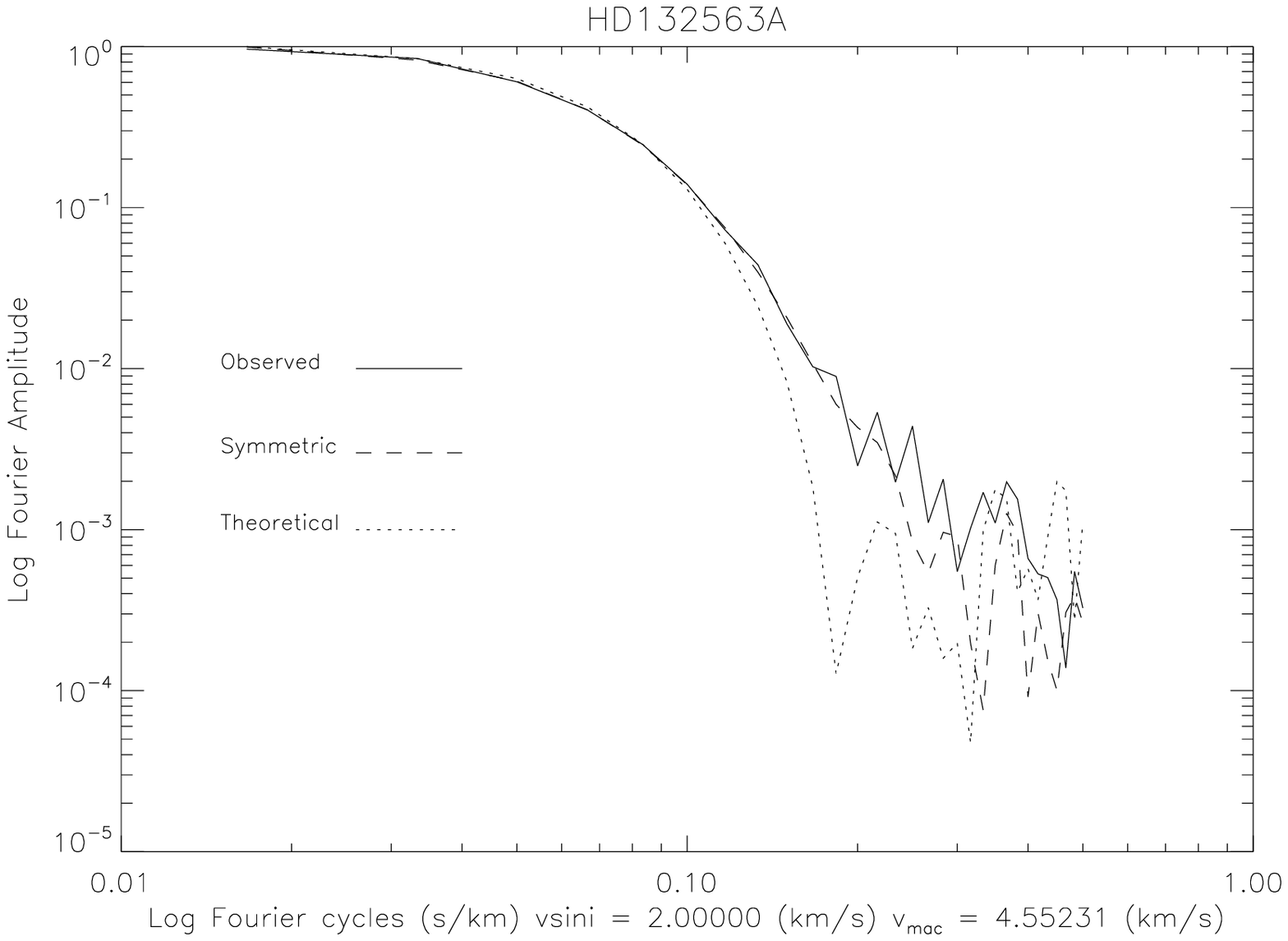}
    \includegraphics[width=9cm]{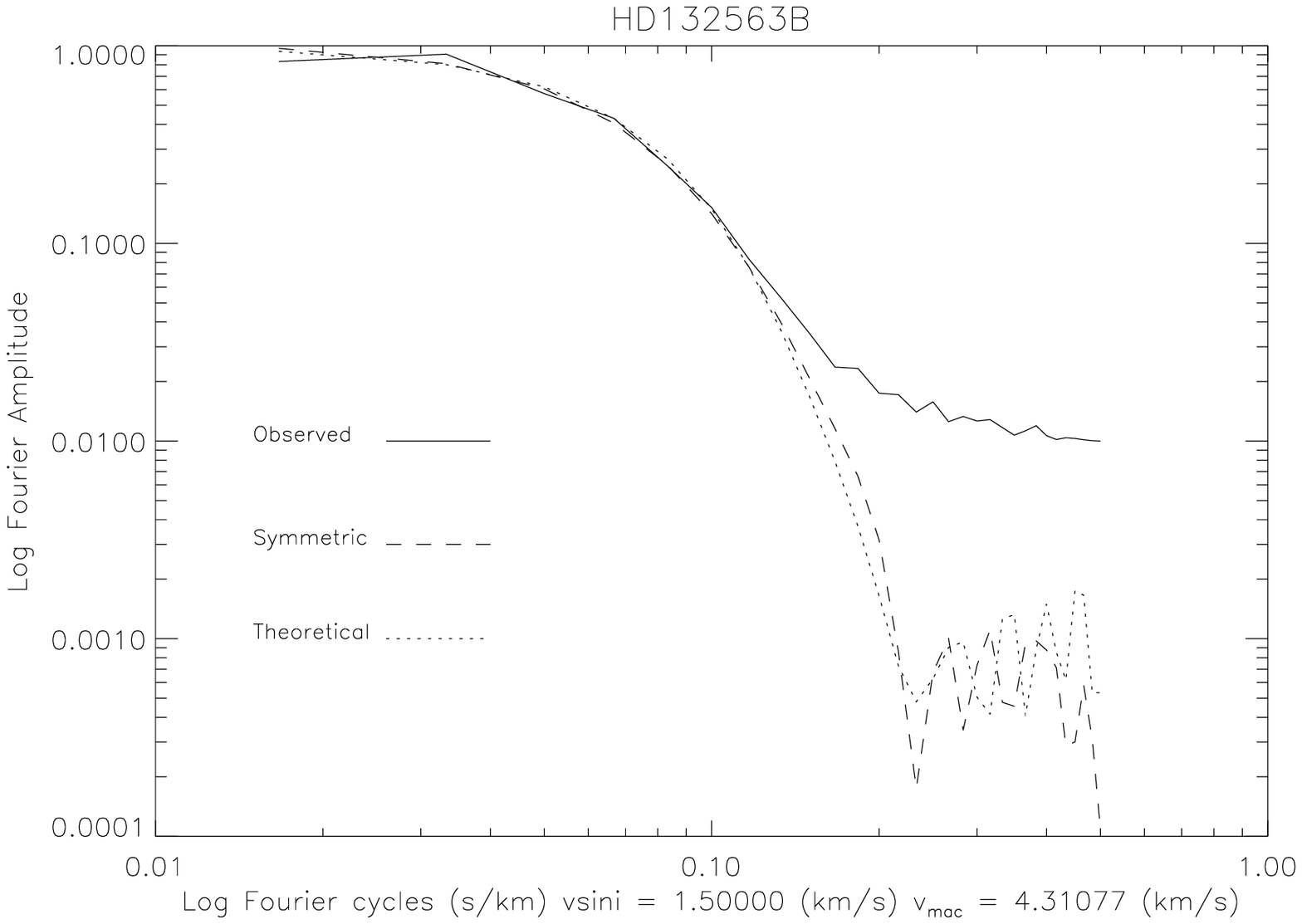}
      \caption{The FFT of HD132563 A and B (upper and lower panels) for three absorption profiles: 
       the solid line for the observed profiles, the dashed line for the symmetric profiles where the FFT 
       appears less noisy, and the dotted line for the calculated (model) profiles.}        
         \label{f:vsini}
   \end{figure}

A direct measurement of magnetic activity level would be very useful for interpreting  the
RV variations in terms of Keplerian motion or activity jitter. Unfortunately, it is not possible to 
simultaneously observe the region that us rich in iodine lines and the CaII H\&K chromospheric lines using SARG.
An indirect estimate in the chromospheric emission can be derived from the measurement of
UV excess in GALEX photometry.
Adopting the joined B and V photometry from Hipparcos 
and the FUV GALEX magnitude of $19.27\pm0.08$,
we get  an UV excess of 
-0.94 mag over the photospheric value, when using the
calibration by Smith \& Redenbaugh (\cite{smith10}). Such an excess is what is expected for
a star with the color of HD132563 and with $\log R_{HK}$ about -4.6\footnote{An alternative explanation
for the small UV excess is represented by a hot white dwarf in the system, 
possibly the spectroscopic companion of HD132563A. However, there are several reasons, 
discussed in Sect.~\ref{s:hd132563a}, to
consider more likely that the spectroscopic companion is an M or K dwarf rather than a white dwarf.}
 (see Fig.~4 in 
Smith \& Redenbaugh \cite{smith10}). The corresponding age using the calibration of
Mamajek \& Hillenbrand (\cite{mamajek08}) is 1.2 Gyr, with a large uncertainty.

The X-ray emission of the object was not detected by ROSAT (Voges et al.~\cite{rosatfaint}). The upper limit
of $\log L_X < 28.6$ (assuming equal X-ray luminosity for the two components and neglecting
the tertiary component around the primary, see below)
corresponds to an age older than about 1.3 Gyr using the calibration of
Mamajek \& Hillenbrand (\cite{mamajek08}).
The star also does not have large photometric variability (0.016 mag dispersion
from Hipparcos measurements, which refer to the joined A+B components and is flagged
as ``duplicity induced variability'', so that the intrinsic variability is likely 
below this value).

\subsection{Absolute radial velocity and Kinematics}
\label{s:kin}

The absolute radial velocity of the components obtained through cross-correlation
with templates of stars from the list of Nidever et al.~(\cite{nidever}) results of
$-6.68\pm0.20$ km/s for HD132563 A and $-6.05\pm0.2$ km/s for B respectively. The systemic
velocity to be used for the kinematic analysis cannot be determined with
accuracy because  of the unknown mass and orbit of the spectroscopic companion
around HD132563A. From plausible values derived in Sect.~\ref{s:montecarlo}, we
adopt a RV of $-6.5\pm1.0$~km/s.
The space velocities, taking trigonometric parallax and proper motion from
van Leeuwen (\cite{newhip}) into account, result in 
U,V,W=$+8.8\pm1.1$, $-39.3\pm4.9$, and $13.5\pm2.8$~km/s. 
The UVW velocities are well outside the kinematic space populated
by young stars (Montes et al.~\cite{montes01}).
The galactic orbit was derived following the procedure described in  
Barbieri \& Gratton (\cite{barbieri}, which assumes the Galactic gravitational potential
by Allen \& Santillan \cite{allen91}).
The numerical integration of the galactic motion yields the following parameters:
$r_{min }=6.30\pm0.31$~kpc, $r_{max}=8.594\pm0.024$~kpc, $z_{max}=0.277\pm0.039$~kpc,
and $e=0.154\pm0.023$. The kinematic properties then indicate an age similar to the Sun
and possibly even slightly older.

\subsection{Stellar age}
\label{s:age}

The isochrone age of $5.5\pm1.5$ Gyr is fully consistent with the kinematic properties of the
system, the slow projected rotational velocity, and the X-ray non-detection.
Lithium and chromospheric activity instead favor a younger age of about 1-3 Gyr.
Considering the lower error in the isochrone fitting, we adopt 5 Gyr as the  most probable stellar age 
 (Table \ref{t:age}).

\begin{table}
   \caption[]{Summary of age estimates of HD132563 components.}
     \label{t:age}
       \centering
       \begin{tabular}{lc}
         \hline
         \noalign{\smallskip}
         Method  &  Age   \\
                 & Gyr     \\
         \noalign{\smallskip}
         \hline
         \noalign{\smallskip}

    Isochrone fitting           & 5.5$\pm$1.5     \\  
    Chromospheric emission     & $\sim 1.2$        \\
    X-ray emission              & $>1.3$           \\
    Lithium                     & 1-4            \\
    Rotation                    & 2-6            \\ 
    Kinematic                   & 3-6            \\
\hline
    Adopted                     & 5          \\

         \noalign{\smallskip}
         \hline
      \end{tabular}%

\end{table}

\section{A giant planet around HD132563B}
\label{s:hd132563b}

\subsection{Radial velocity variations of HD132563B}
\label{s:rvb}

\begin{figure}
   \includegraphics[width=9cm]{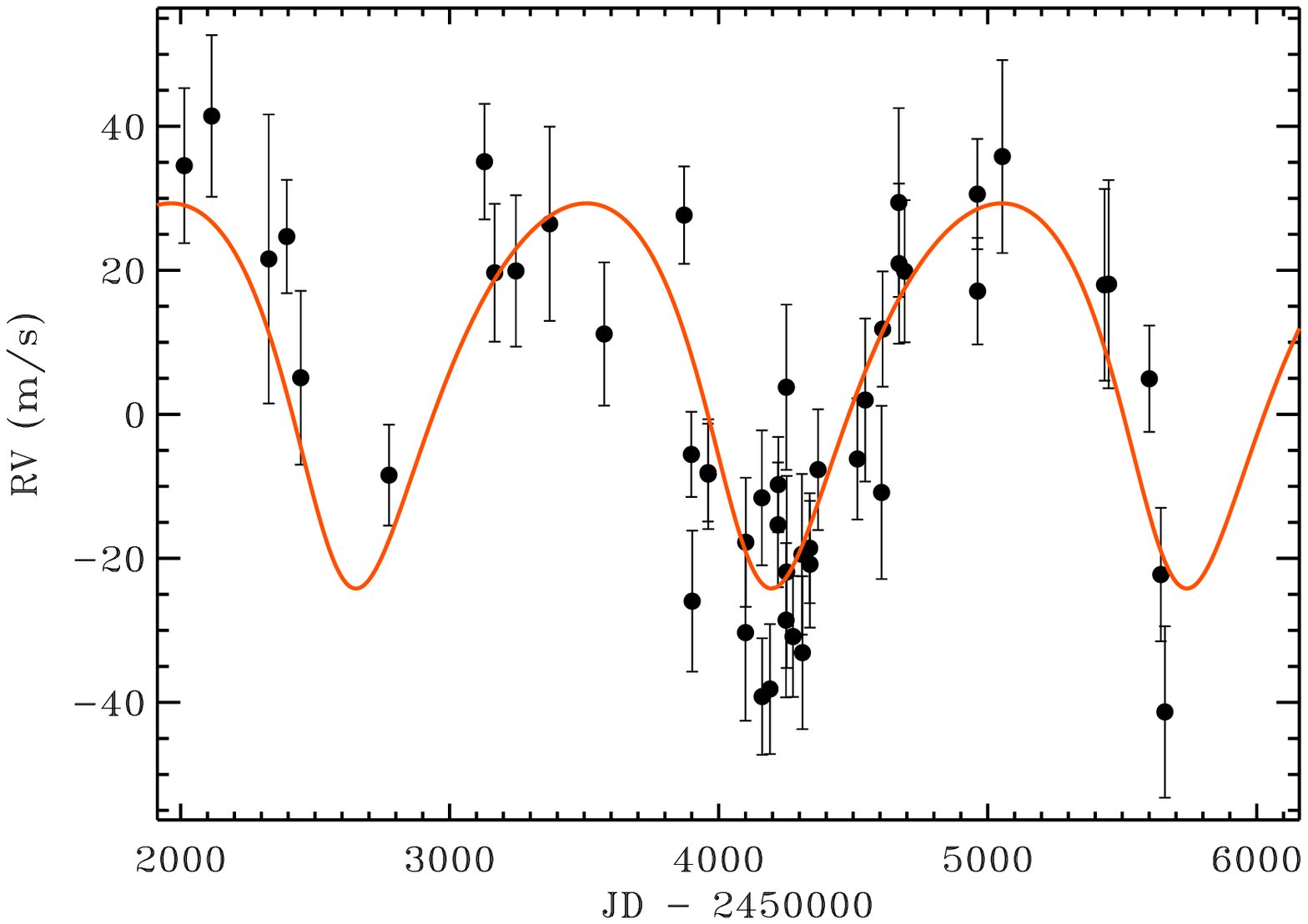}
   \includegraphics[width=9cm]{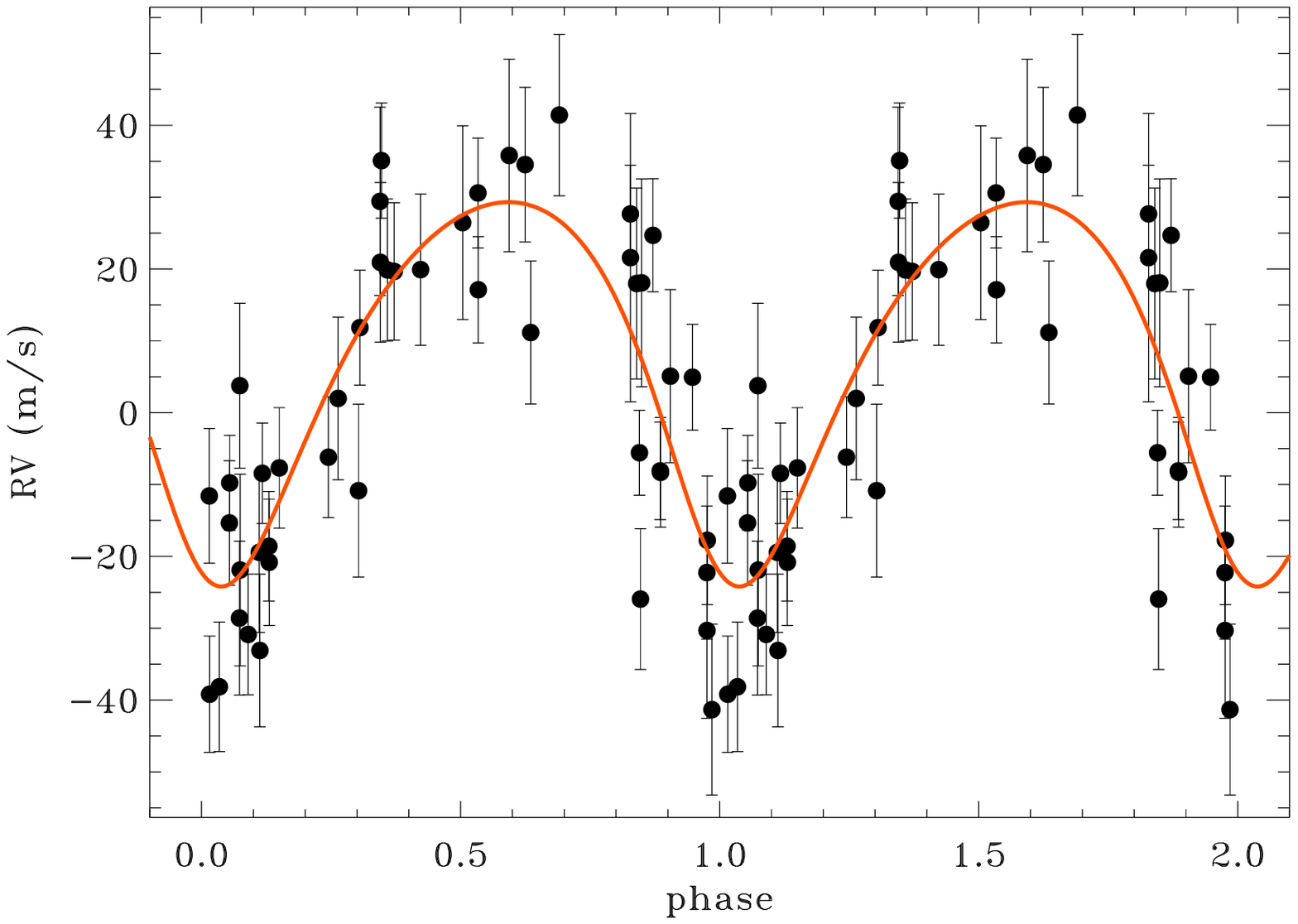}
   \includegraphics[width=9cm]{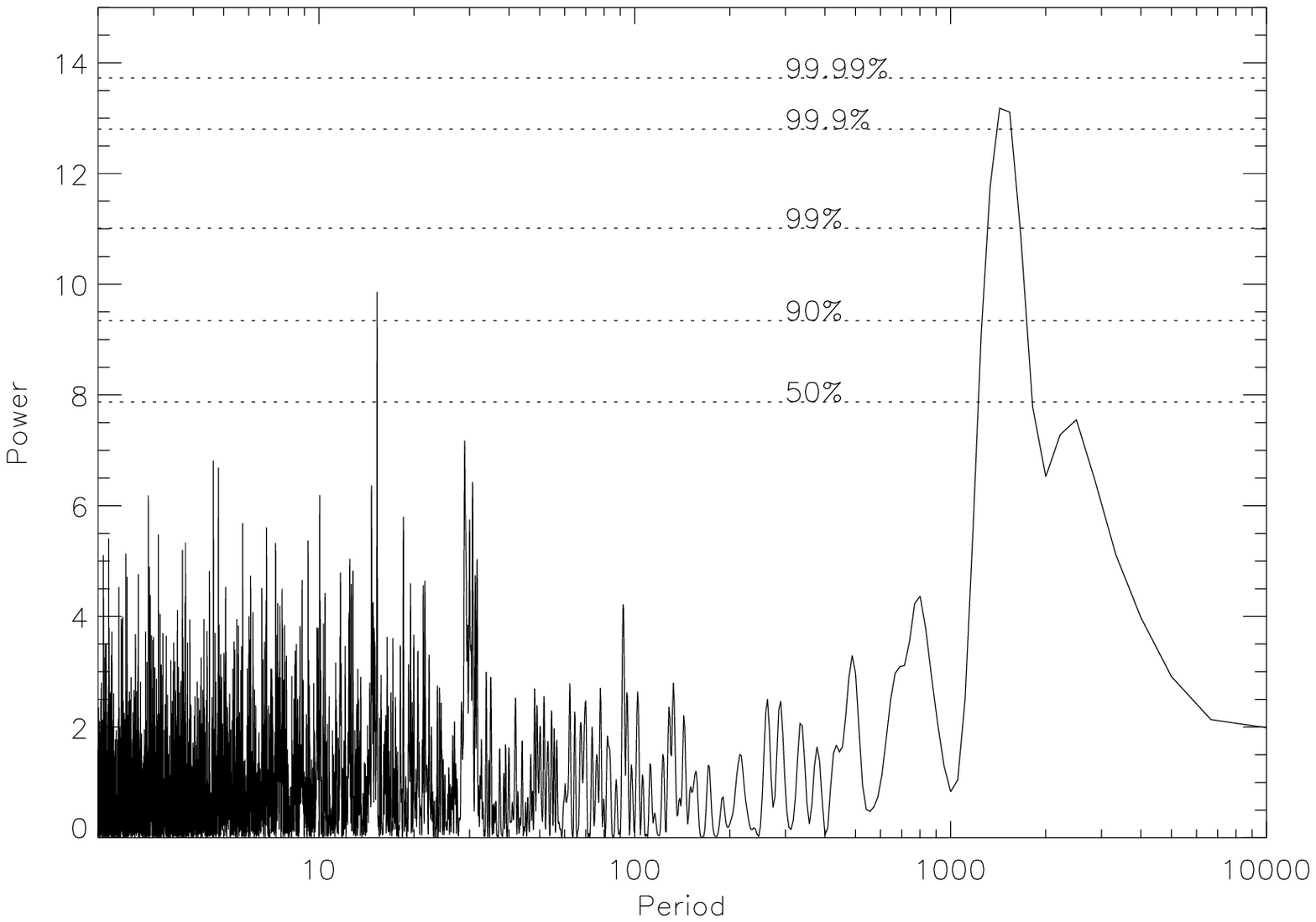}
      \caption{Upper panel: radial velocities of HD 132563 B, overplotted is
               the Keplerian best fit. Central panel: radial velocities phased to
               the best-fit period. Lower panel: Lomb-Scargle periodogram of 
               the radial velocities. Overplotted are the various confidence levels as
               derived from the bootstrap analysis.}        
         \label{f:rv_hd132563b}
   \end{figure}

The RVs of HD132563B (Table \ref{t:rvb}) show significant variations (r.m.s. 23.4 m/s) compared to
internal errors (10.3 m/s).
Internal errors are larger than those for the typical targets of the SARG planet search because
the object is at the faint end of our sample.
The Lomb-Scargle periodogram (Scargle \cite{scargle}) shows a well-defined peak at about 1500 d. 
Bootstrap simulations performed by scrambling the RVs  yield a false alarm probability of
5/10000 (Fig.~\ref{f:rv_hd132563b}). This is low enough to provide confidence that the detected RV periodicity is real.

\subsection{Origin of RV variations}
\label{s:origin}

Before inferring the Keplerian nature of the observed variations we have to exclude
alternative possibilities such as magnetic activity.
An activity jitter of 20 m/s is usually observed in
stars of similar color with $\log R'_{HK} \sim -4.4 \div -4.5 $ and age of about 400 Myr.
As discussed in Sect.~\ref{s:star}, we do not have any indication of such a 
strong magnetic activity of the star or such a young age.  
Furthermore the observed RV period is much longer than the expected rotational period
of a solar type star.

\begin{figure}
   \includegraphics[width=9cm]{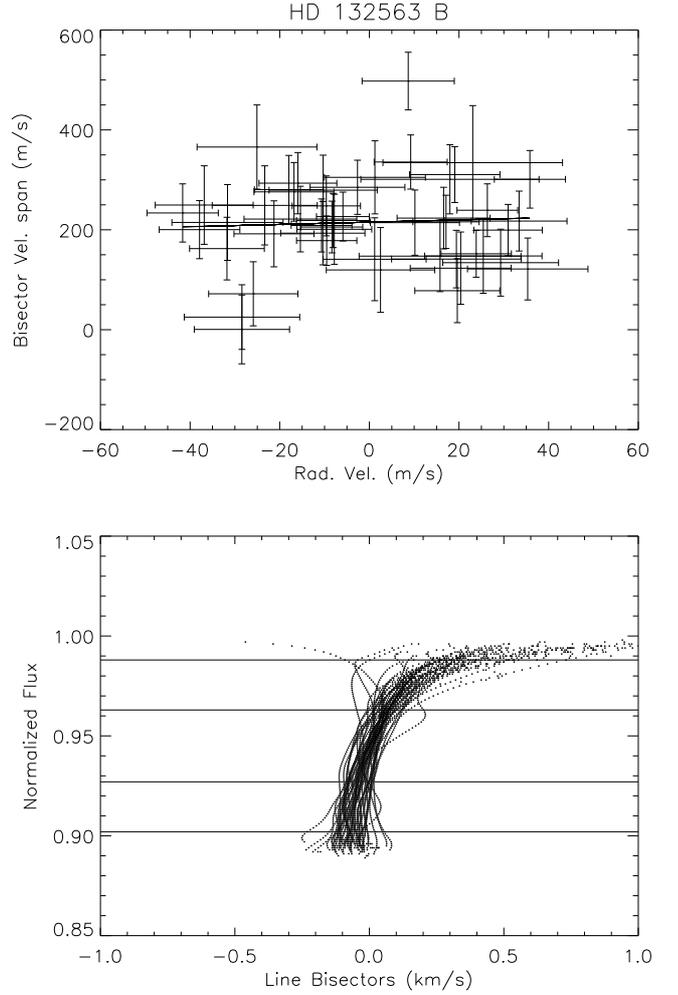}
      \caption{Upper panel: bisector velocity span (BVS) vs radial velocity for the spectra of HD132563B,
               showing no correlation.
               Lower panel: Line bisectors of individual spectra of HD132563B. The parts of the line 
               profile adopted for the derivation of BVS are shown as horizontal lines.}        
         \label{f:hd132563b_bvs}
   \end{figure}

Finally, activity-induced variations are usually accompanied by
changes in line profiles correlated with those of radial velocities, 
even if this diagnostic is less senstive for slowly rotating stars such as
HD132563B; see Santos et al.~(\cite{santos03}) and Desort et al.~(\cite{desort07}).
To this aim, we performed the analysis of line  bisectors  using the technique described in 
Martinez Fiorenzano et al.~(\cite{aldo}).
As shown in Fig~\ref{f:hd132563b_bvs}, there is no significant correlation between RV and bisector 
velocity span (BVS).

Another potential source of spurious RV variations in multiple systems is represented 
by contamination of the spectra by the light of the other components. 
This is not of concern in the present case considering the projected
separation of about 4.1 arcsec. A few spectra taken with a seeing of about 2 arcsec show no
indication of significant contamination neither as outliers in the RV curve nor in
the line profile and there is no correlation between RV and seeing.

We then conclude that the most likely explanation for the observed RV variation is the
presence of a low-mass object, \object{HD132563Bb}, whose projected mass is clearly in the planetary regime.

\subsection{Orbital parameters}
\label{s:orbit}

Orbital parameters to characterize the detected RV variations of HD132563B were
determined by using an our own code, which is based on a Levenberg-Marquardt least-squares fit
of the RVs. An additional error of 7 m/s (representing activity jitter and possible unrecognized systematic errors) 
was added in quadrature to internal errors when performing the fit.
The best-fit parameters are listed in Table \ref{t:fit} and the best-fit solution 
plotted in Fig.~\ref{f:rv_hd132563b}. 
The errors in the orbital parameters were derived by simulating synthetic datasets 
that take the adopted jitter and performing orbital fitting into account for each
fake RV series. The r.m.s. of residuals
from the fit are 12.7 m/s, slightly larger than the internal errors.
A circular orbit fitting yields slighly larger residuals (13.5~m/s), indicating 
that the value eccentricity is rather uncertain.
From the stellar mass $1.010~M_{\odot}$, we derive a projected mass of 1.49 $M_{J}$
and a semimajor axis of 2.62 AU for HD132563Bb.

\begin{table}
   \caption[]{Orbital parameters and results of fitting for RV of HD132563B.}
     \label{t:fit}
       \centering

 \begin{tabular}{lc}
         \hline
         \noalign{\smallskip}
         Parameter  &  Value    \\      
         \noalign{\smallskip}
         \hline
         \noalign{\smallskip}
Period (d)                   & 1544$\pm$34  \\
K (m/s)                      & 26.7$\pm$2.2    \\
e                            & 0.22$\pm$0.09    \\
$\omega$ (deg)               & 158$\pm$35   \\
T0                           & 2593$\pm$148 \\
$m \sin i$ ($M_{J}$)         & 1.49$\pm$0.09    \\
a (AU)                       & 2.62$\pm$0.04   \\
r.m.s. res  (m/s)               & 12.7          \\
      \noalign{\smallskip}
         \hline
      \end{tabular}

\end{table}

\begin{figure}
   \includegraphics[width=9cm]{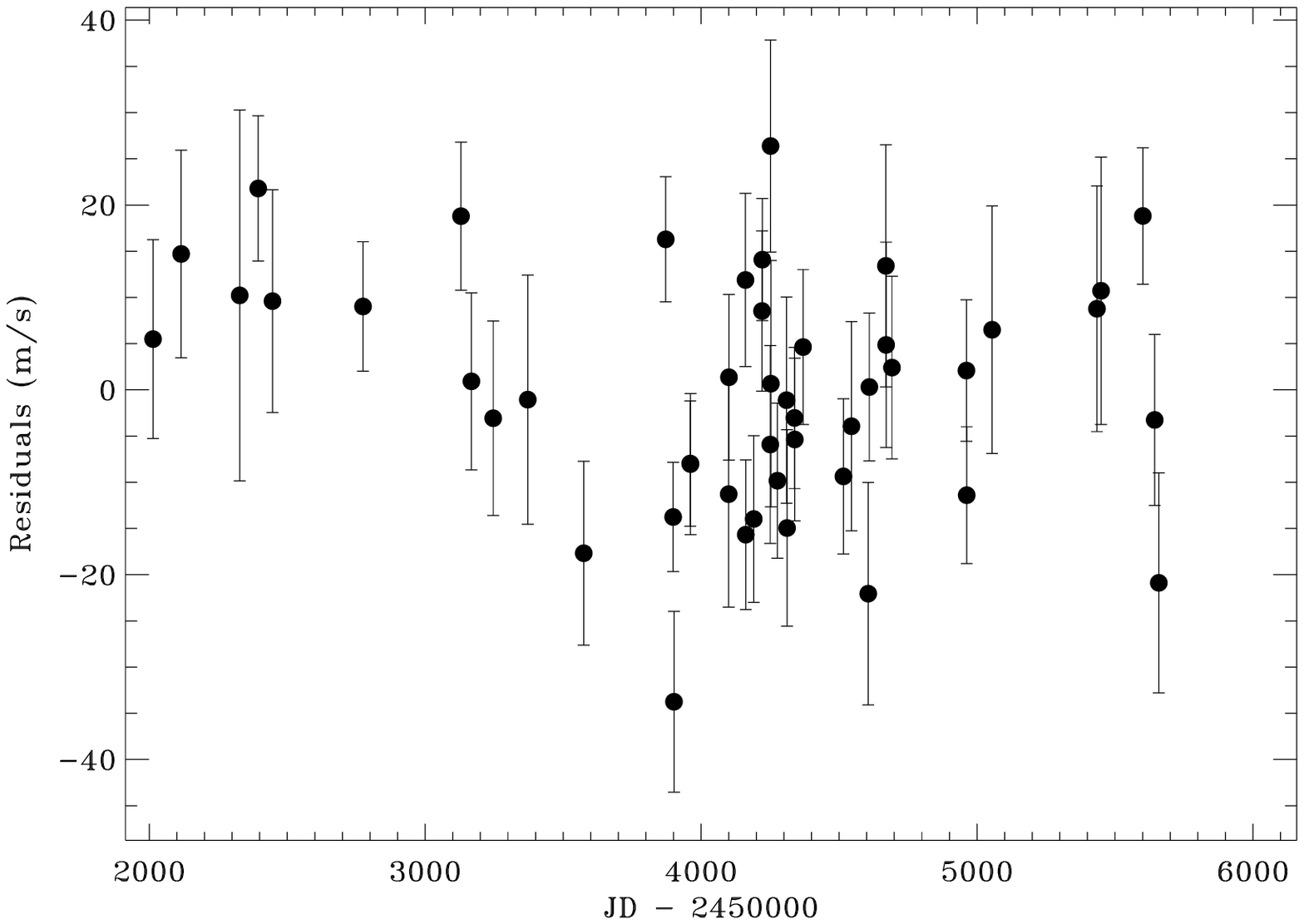}
   \includegraphics[width=9cm]{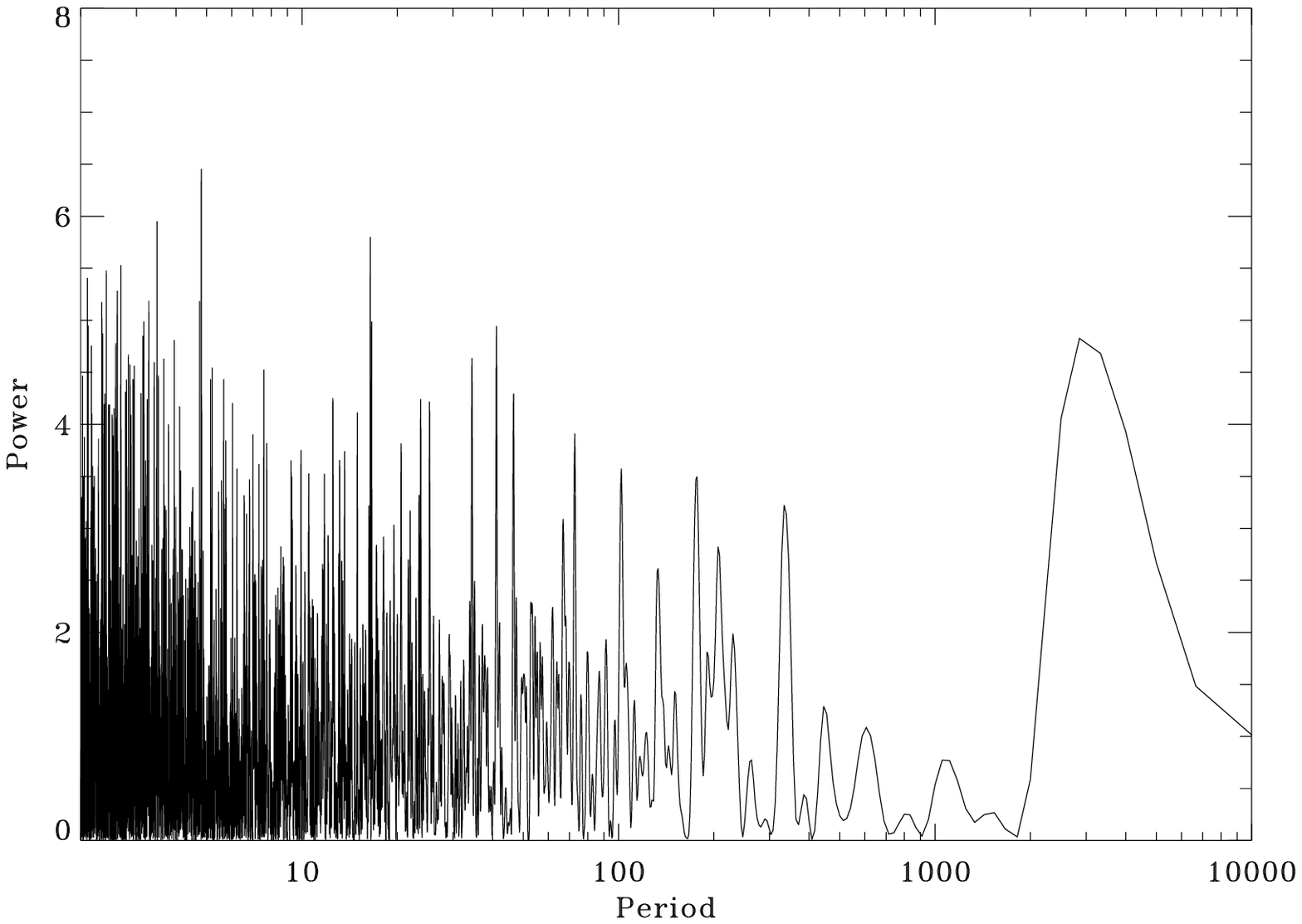}
      \caption{Upper panel: Residuals from best fit orbit
               Lower panel: Lomb-Scargle periodogram of residuals from formal Keplerian best fit of the
               radial velocities of HD132563B.}        
         \label{f:res_hd132563b}
   \end{figure}

Analysis of residuals from the best-fit orbit does not reveal any significant 
periodicities (Fig~\ref{f:res_hd132563b}).
There is a possible long-period modulation that is not highly significant
but that calls for the continuation of the RV monitoring to look whether it is
real and, in this case, if it might be due to an additional planet in outer orbit.
The highest peak in the periodogram of residuals in the range 10-30 days 
is at 16.4 days, which is different from the second highest peak in periodogram 
of the originl RVs (15.3 days). Such a periodicity might be related to stellar rotation
but can just come from noise fluctuations.

\section{A stellar companion in highly eccentric orbit around HD132563A}
\label{s:hd132563a}

\subsection{Radial velocity variations of HD132563A}
\label{s:rva}

\begin{figure}
   \includegraphics[width=9cm]{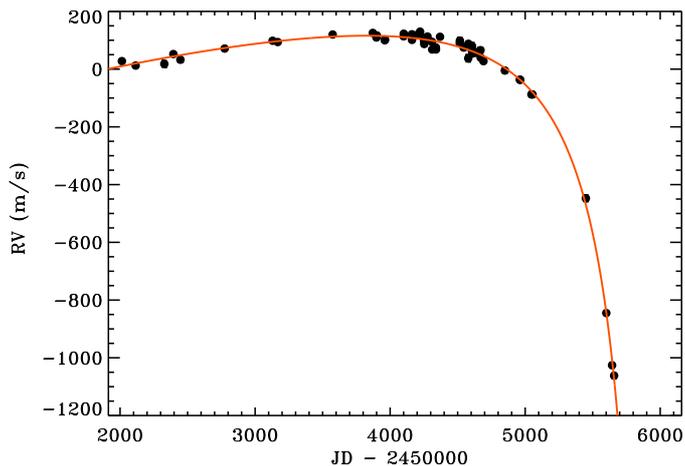}
      \caption{Radial velocities of HD~132563A. Overplotted
               the Keplerian best fit.}        
         \label{f:rv_hd132563a}
   \end{figure}

The RV of HD132563A showed a nearly linear trend for a few years. The RV maximum 
was reached in 2006. After this, an RV decrease with increasingly steep slope
was observed (Fig.~\ref{f:rv_hd132563a}). The peak-to-valley RV variations are now more than 1 km/s, 
placing the companion to HD132563A well outside the planetary mass regime\footnote{Following
standard notation, we identify the primary of the spectroscopic pair as \object{HD132563Aa}, contributing to a large
fraction of the observed flux, and its spectroscopic companion as \object{HD132563Ab}.}.
The orbit is highly eccentric.
Residuals from the formal best-fit orbit do not show any significant periodicities.
The dispersion of the residuals is 14.1 m/s, which are significantly larger
than the internal errors (8.5 m/s), leaving room for a jitter of about 11 m/s\footnote{An additional error in the RVs might 
also be expected if the
contribution of HD132563Ab to the integrated spectrum is not negligible. In this case,
the velocity difference between the spectroscopic components changes with time, and the template,
which is unique for all the observations, might not be fully representative of the true stellar spectrum
at the various epochs, causing some degradation in the modeling  of the composite iodine+star spectrum.}.

Because only a portion of the orbit of HD 132563 A was observed,
a wide variety of orbital solutions and companion masses  are
compatible with the RV data (Sect.~\ref{s:montecarlo})
However, additional, available data allow us to further constrain
the possible parameters of the companion.
They are described below individually, and conclusions on the
possible mass and orbit are presented in Sect.~\ref{s:summary_a}.

\subsection{Montecarlo simulation}
\label{s:montecarlo}

\begin{figure*}
\includegraphics{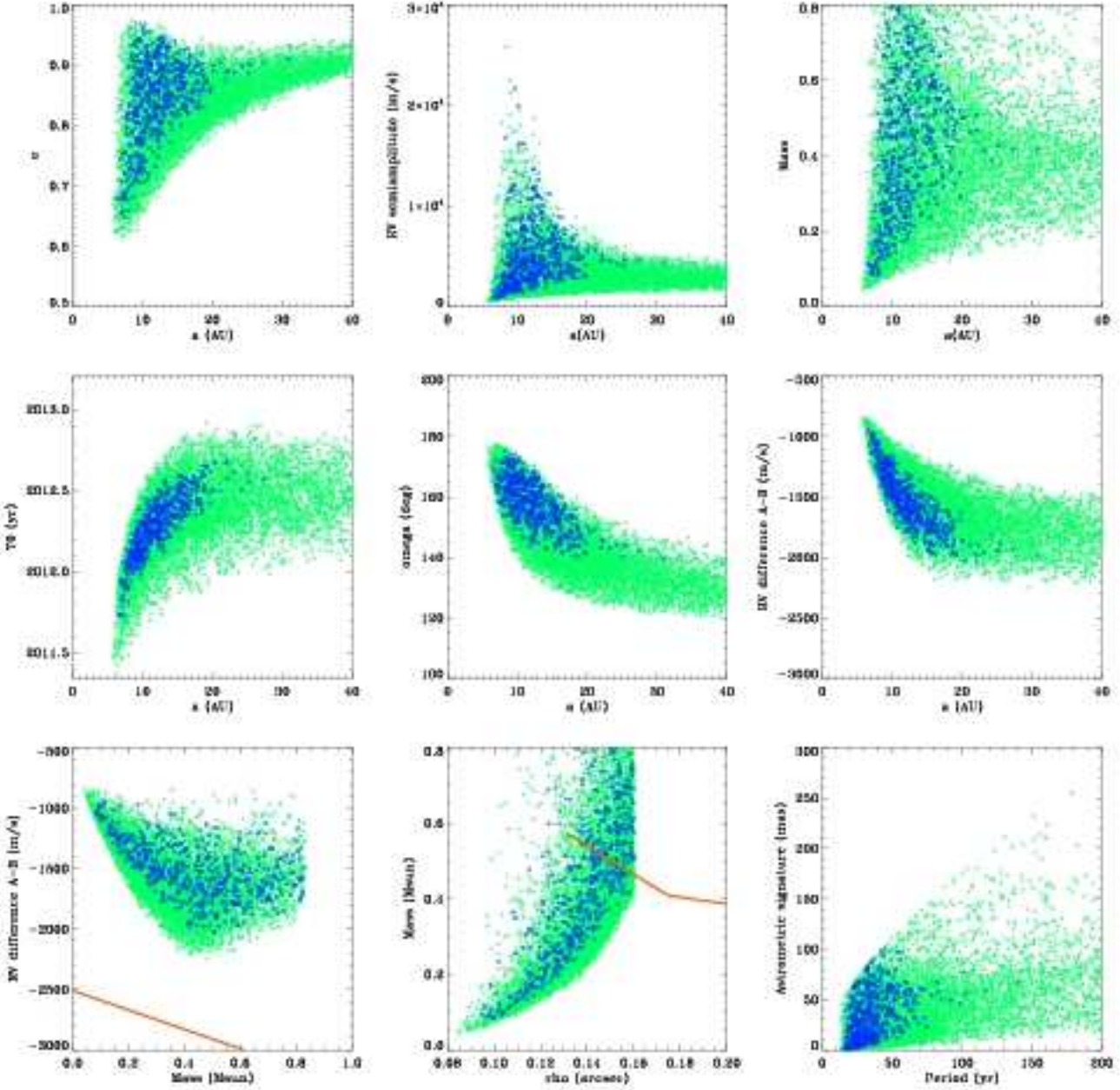}
\caption{Results of the MonteCarlo simulation, showing the possible parameters that are
compatible with the observed RV curve of HD132563A. Starting from top left: eccentricity, 
RV semiamplitude, companion mass, time of periastron passage, longitude of periastron,   
RV difference between the components vs semimajor axis; RV difference between the components vs 
companion mass; companion mass vs projected separation (at epoch 2007.548); astrometric amplitude
vs orbital period.
Blue circles: solution with r.m.s. of residuals $<15.7$ m/s; Green circles (light gray in BW):
solution with r.m.s. of residuals $15.7-18.5$ m/s.
The red solid line in the mass vs RV difference panel is the limit that allows the wide
pair to be bound. The red solid line in the companion mass vs projected separation panel
shows AdOpt@TNG detection limits.}
   \label{f:simul}
   \end{figure*}

To take the various constraints on the mass and orbital parameters of the
spectroscopic companion into account we performed a dedicated MonteCarlo simulation.
The procedure first generates a set of companion masses and all the orbital parameters.
We established an upper limit on the semimajor axis at 40 AU (corresponding to a period of about 200 yr).
Wider orbits  appear unlileky because typically the inner pair in triple systems has an orbit
that is much closer than the outer component, well within the dynamical stability region and 
the probability of catching the phase of steep RV variation
at periastron passage in an RV monitoring of about ten years is lower than 5\%.
Furthermore, the relative astrometry does not indicate any periodicities
that are a few times the time baseline of the observations, which would be expected because
the astrometric amplitude is larger at longer periods for given companion mass, and
best-fit solutions in our simulation are confined to shorter periods (see below).
We also fix the upper limit of the companion mass at $0.8~M_{\odot}$, from the lack
of signatures in our spectra (Sect.~\ref{s:spectra}).
From each generated input set, we derived the corresponding RV variations at 
the time of SARG observations, the projected separation
and luminosity contrast at the time of AdOpt observations, and the astrometric amplitude 
for comparison with the observational data.
This allowed us to identify the combination of parameters that produces an RV curve
compatible with the observed one (r.m.s. of residuals $<$ 18.5 m/s, 2$\sigma$ limit as compared to
the residuals from the formal best-fit orbit using F test; Fig.~\ref{f:rv_hd132563a}).
Figure~\ref{f:simul} shows the results of the simulations: there is a broad range of orbital parameters
that are compatible with the RV signature, which becomes narrower when considering
only the best-fit solutions (1$\sigma$ limit; r.m.s. of residuals $<15.7$~m/s). 
The orbit results highly eccentric ($e>0.65$), with periastron passage expected within the end of 2012.
Table \ref{t:orbit_a} lists the most probable orbital parameters as resulting from our simulations.

\begin{table}
   \caption[]{Possible orbital parameters of HD132563A, considering the orbital solutions that give
   r.m.s of residuals $<15.7$ m/s.}
     \label{t:orbit_a}
       \centering
       \begin{tabular}{lccc}
         \hline
         \noalign{\smallskip}
     Parameter  &  Mean & Median & r.m.s.  \\
         \noalign{\smallskip}
         \hline
         \noalign{\smallskip}
a  (AU)                &   14.8  &     12.9  &     6.3  \\
P  (yr)                &   47   &     37   &    30 \\
e                      & 0.860   &   0.874   &  0.065 \\
K (m/s)                &  5460   &    4887   &    2995 \\
$m \sin i (M_{\odot}$) &  0.40   &      0.38 &     0.18 \\
mass ($M_{\odot}$)     &  0.49   &     0.51  &    0.20  \\
$\rho$ (arcsec)        &  0.145  &    0.146   &  0.018  \\
$\omega$ (deg)         &   160.2 &       160.0 &    5.9  \\
T0   (yr)              & 2012.34 &    2012.34   &  0.20  \\
$\gamma$ (m/s)         &   -1625 &      -1613 &      288  \\
astrom. sign. (mas)    &      51 &       40 &      43  \\

         \noalign{\smallskip}
         \hline
      \end{tabular}

\end{table}

\subsection{Spectral properties and line bisectors}
\label{s:spectra}

\begin{figure}
   \includegraphics[width=9cm]{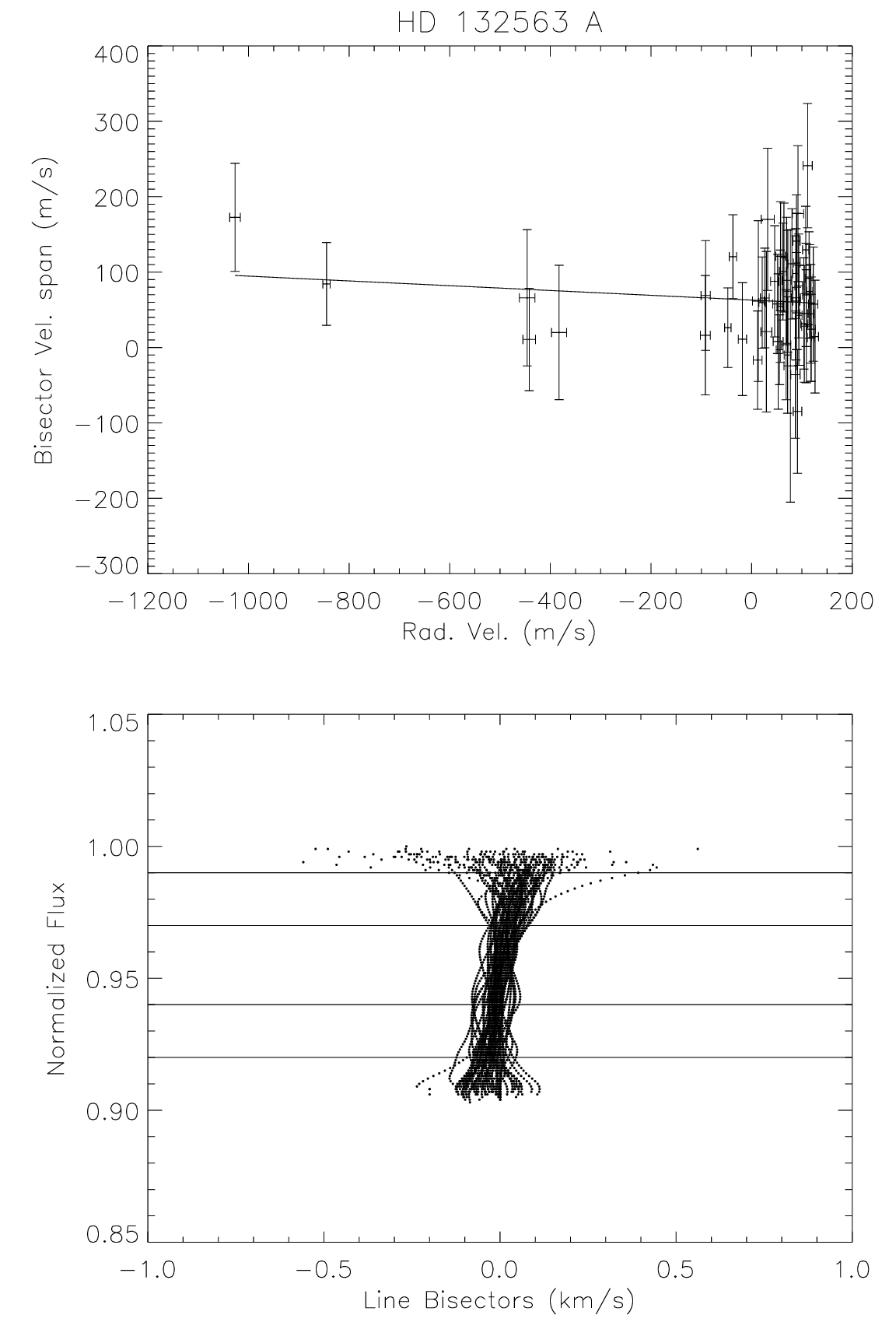}
      \caption{Upper panel: Bisector velocity span (BVS) vs radial velocity for the spectra of HD132563A.
               No correlation is present.
               Lower panel: Line bisectors of individual spectra of HD132563A. The parts of the line 
               profile adopted for deriving  BVS are shown as horizontal lines.}        
         \label{f:hd132563a_bvs}
   \end{figure}

We analyzed the spectra of HD132563A and the resulting line profiles looking for
signatures of the spectroscopic companion on the spectra (Fig.~\ref{f:hd132563a_bvs}).
There is no significant correlation between RV and BVS and the individual line profiles typically
appear clean without indicating the spectral signatures of the spectroscopic companion.
As a rough guess, we estimate that secondaries contributing more than 10\% of the light in the 
{\em V} band 
are excluded by the line profile analysis.
The absolute shape of the line bisector, less 'C shaped' than HD132563B, might indicate
of a small contamination of the line profile at negative RV (where we expect the contribution
of the spectroscopic companion), at the level of a few \%.

\subsection{Magnitude difference between the components}
\label{s:kexcess}

As shown in Sect.~\ref{s:isoc}, a fairly good isochrone fitting is obtained
by exploiting the small errors on the V magnitude and temperature differences.
The K mag difference, $0.467\pm0.010$, is not compatible with what is expected 
from the best fit isochrone.
The magnitude difference is larger by about 0.09 mag. 

Because we know that there is an additional component in the system, HD132563Aa,
we derive its properties assuming that it is the cause of the K band excess
and that the other two components follow the best-fit isochrone.
We found $M_{K}$=2.76,  3.14, and 5.45 for HD132563Aa, HD132563B, and HD132563Ab, respectively.
Using the mass-luminosity relation by Delfosse et al.~(\cite{delfosse00}), the corresponding
mass of the spectroscopic companion would be $0.56~M_{\odot}$.
The relative contribution of HD132563Ab in V band would be smaller, about 0.7\%, with little
effect on the isochrone fitting.

\subsection{Direct imaging}
\label{s:ao}

No companions close to either HD132563 A or B were detected in our images taken with
AdOpt@TNG (Sect.~\ref{s:adopt}). The procedure of subtracting the PSF of the other component 
leaves residuals that cannot be eliminated by optimizing the position shift between the components.
This indicates a slightly different PSF between the two components.  
A companion contributing less than 10\% at a very small projected separation might produce such feature, which
might also, however, be related to instrumental effects.

\begin{figure}
   \includegraphics[width=9cm]{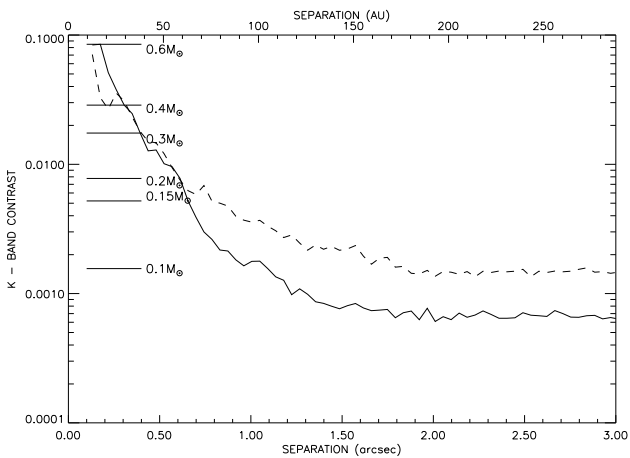}
      \caption{K band detection limits for observations taken with AdOpt@TNG in July 2007
(dashed line) and August 2008 (continuous line). A few individual values of masses 
corresponding to contrast values are also shown.}        
         \label{f:limitsadopt}
   \end{figure}

To estimate detection limits, we considered the dispersion of fluxes of pixels 
in circular annuli around the calculated position of the star, with steps of one pixel.
The detection limits were set at five times the standard deviation in each annulus.
During these calculations the pixels at less than 30 pixels from the other component 
were masked.
In Fig.~\ref{f:limitsadopt} we display the contrast obtained around the A component and the corresponding mass 
detection limits. 
To translate the contrast limit on companion masses, we adopt the $K$ magnitude of the components from 
2MASS (Skrutskie et al.~\cite{2mass})\footnote{2MASS photometry for close pairs of bright stars is rather poor, 
with the error exceeding the nominal ones because of blending.}
and use the mass-luminosity relation by Delfosse et al.~(\cite{delfosse00}) for companions with $M_{K}<9$ 
and Chabrier et al.~(\cite{chabrier}) models at fainter magnitudes (using the adopted stellar age).

The expected projected separation at the epoch of our observations is between 0.09
and 0.18 arcsec (Fig.~\ref{f:simul})
at 2007.548 and smaller by about 10-15\% one year later, as the pair is approaching periastron passage.
The contrast limit derived at such separations corresponds to about $0.6~M_{\odot}$\footnote{Detection limits
are uncertain at small projected separations, where there are few points in the circular annuli.
Furthermore, a moderately bright companion can affect the centering procedure.}. 
Detection limits at separations larger than 1.5 arcsec (background-limited region)
is close to the stellar/substellar boundary.

The mentioned  mass limits apply to main sequence stars. White dwarfs are faint in $K$ band
(Bergeron et al.~\cite{bergeron01}) and are not excluded by our direct imaging data. 
However, a white dwarf appears unlikely as close companion. 
HD 132563A is a close enough binary that the (current) primary HD132563Aa should have 
accreted a significant amount of material with the signature of AGB nucleosyhthesis,
becoming a barium dwarf, if its companion HD132563Ab has evolved through the AGB phase
and is now a white dwarf (see McClure et al.~\cite{mcclure80}; Frankowski \& Jorissen \cite{frankowsky07}). 
However,  the barium abundance of HD132563A is identical within 0.01 dex to that of HD132563B.
The small FUV excess of the system can be explained by a moderate
chromospheric activity of the stars.

\subsection{Relative astrometry}
\label{s:astrometry}

\begin{figure}
\includegraphics[width=8.5cm]{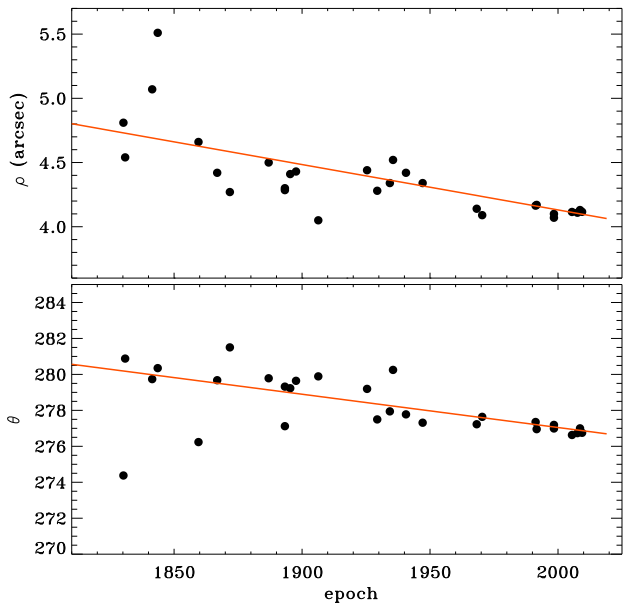}
\caption{Relative motion of HD 132563. 
         Continuous line is the linear fitting to the data.}
   \label{f:rhotheta}
\end{figure}

The relative astrometry between the components of HD132563 can be used
to further constrain the properties of the companion
around HD 132563A.
Available astrometric measurements in the Washington Double Star Catalog 
(Mason et al.~\cite{wds}, version 4 Apr. 2011, kindly provided to us by B.~Mason) are shown in
Fig.~\ref{f:rhotheta}.
The astrometric time series over a span of about 180 years clearly show
the signature of the orbital motion. As no curvature is detected, we assume
this is due to the motion of the wide pair and not to the spectroscopic companion
to HD132563A (otherwise the period of the companion would be very long, making the probability
of catching the object at periastron passage very low).
The linear best fit yields slopes of
$d \rho / dt = -0.00353 \pm 0.00036 $ arcsec/yr and $d \theta / dt = -0.0185 \pm 0.0033 $ deg/yr.

\begin{figure}
\includegraphics[width=8.5cm]{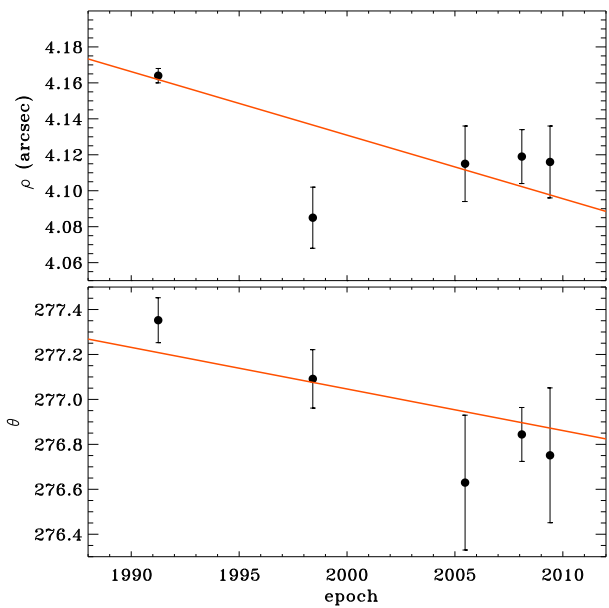}
\caption{Recent high-quality relative astrometry of HD132563 system. 
         Continuous line is the linear fitting to the data shown in 
         Fig.~\ref{f:rhotheta}. The measurements from Douglass et al.~(\cite{douglass00}) 
         and AdOpt@TNG were averaged.}
   \label{f:rhothetabest}
\end{figure}

Figure \ref{f:rhothetabest} shows the relative astrometry of HD132563
in the last 20 years considering only high-quality measurements:
Hipparcos, speckle interferometry (Douglass et al.~\cite{douglass00}; 
Scardia et al.~\cite{scardia07}; Losse \cite{losse10} and our own AdOpt measurements from Table~\ref{t:adopt}).
While the details of the residuals pattern depend on the adopted long-term slope, which
has some uncertainty,  the measurement by Douglass et al.~(\cite{douglass00}) is well outside 
of the linear fit obtained considering all epochs back to 1830.
It is plausible that we are seeing the astrometric impact of  HD132563Ab
on HD132563Aa. If this is the case, a lower limit to the amplitude of the astrometric motion
would be about 50 mas. This is compatible with the astrometric signature of a 
$0.56~M_{\odot}$ companion, as suggested by the K band excess, and the most
probable orbital elements from RV.

\subsection{Radial velocity difference between the components}
\label{s:rvdiff}

As HD132563A is approaching the periastron of the spectroscopic orbit, its RV becomes more negative, and the RV difference
between the components increases. However, for increasing values of the RV difference between
HD 132563 A and B
(because of  a high value of RV amplitude of the spectroscopic variations) it would become no
longer possible to ensure that the two wide components are bound.
We derived such a limit using the approach presented by Hauser \& Marcy (\cite{hauser}).

We first obtained the RV difference between the components of HD132563
by measuring the RV of HD132563B with the stellar template of HD132563A.
We found an RV difference between the mean of the RVs of each component of
$\Delta RV (A-B)= -747\pm35$~m/s.
The mean RV of HD132563B is very close to the center of mass of the star+planet system,
while the  center of mass of HD132563A is unknown because of the still undetermined
spectroscopic orbit.
However, from the orbital solutions compatible with the RV data we can infer that the absolute RV 
difference between the components should be between
$-900$ to $-2100$ m/s, with a dependence on the mass of the spectroscopic companion 
(and semimajor axis, see Fig.~\ref{f:simul}; Sect.~\ref{s:montecarlo}).

From the relative astrometry of the components, assuming that what we observe is
their relative orbital motion, we derived the relative positions and velocities between the 
components on the plane of the sky ($x, y, v_x, v_y$).
These result $x=52$~AU, $y=-397$~AU, $v_x=-814$~m/s, and 
$v_y=1515$~m/s. The RV difference between the components yields the 
velocity along the line of sight ($v_z$). 
The relative position along the line of sight ($z$) remains unknown.

The maximum separation for having a bound pair is 
\begin{equation}
r_{max}=\sqrt{x^2+y^2+z_{max}^2}=\frac{2.0~G~m_{A}~m_{B}}{m_{red} ~v^2}
\end{equation}

\noindent where $m_{A}$ and  $m_{B}$ are the masses of HD132563 A and B, respectively,
$m_{red}$ is the reduced mass, and $v$  the orbital velocity $v= \sqrt{v_x^2+v_y^2+v_z^2}$

The RV differences between A and B that do not allow the system to be bound would be lower
than $-2.67$ and $-3.07$ km/s for masses of HD132563Ab of 0.20 and 0.70~$M_{\odot}$ respectively.
The actual values of the plausible orbital solutions are well above this limit (Fig.~\ref{f:simul}).

\subsection{Conclusion on the characteristics of the spectroscopic companion}
\label{s:summary_a}

The results described above does not allow us to fully and conclusively determine  the mass of
HD132563Ab but are all consistent with a $0.5-0.6~M_{\odot}$ companion.
Such a mass is suggested by the small K band excess and the tentative evidence of astrometric
signature. It is close to the most probable value suggested by the Montecarlo simulation.
Contribution to the integrated flux of HD132563A is expected to be at a $\le1$\% level in the visible, 
with
negligible impact on the various results presented in the paper.
The RV variations strongly constrain the orbital period, eccentricity,
longitude of periastron and epoch of periastron passage. The continuation of the RV monitoring
will allow a firm determination of these parameters and of the projected mass.

\section{Binary orbit}
\label{s:binorbit}

\begin{figure}
\includegraphics[width=8.5cm]{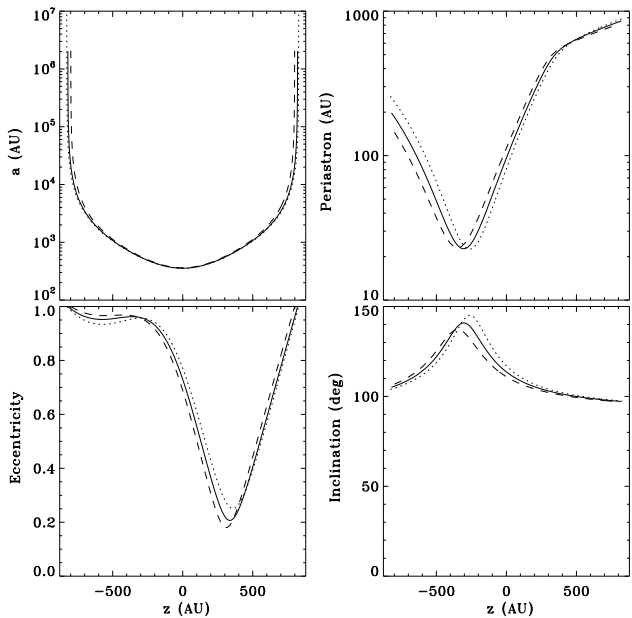}
\caption{Semimajor axis, eccentricity, periastron, and inclination
of the binary orbit as a function of the separation along the line
of sight $z$ for various RV differences between the components
and masses of HD132563Ab.
Continuous line: RV difference -1.4 km/s and mass $0.45~M_{\odot}$.
Dotted line: RV difference -1.2 km/s and mass $0.20~M_{\odot}$.
Dashed line: RV difference -1.6 km/s and mass $0.70~M_{\odot}$.}
   \label{f:binorbit}
   \end{figure}

\begin{figure}
\includegraphics[width=8.5cm]{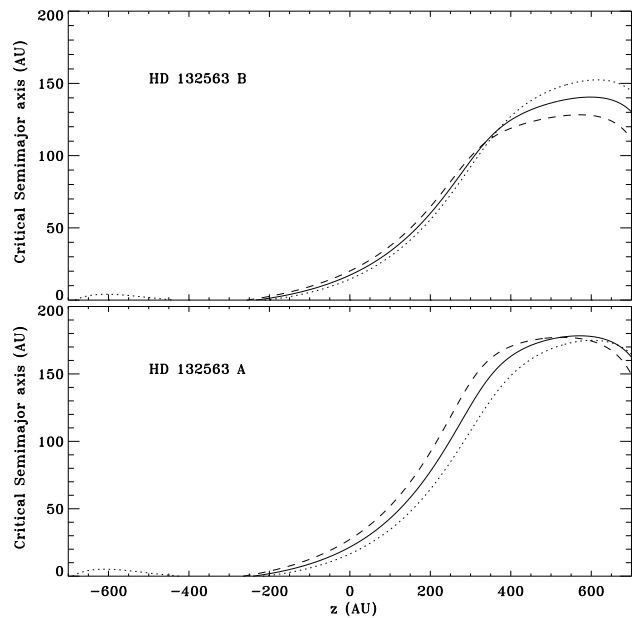}
\caption{Critical semimajor axis for dynamical stability for the various wide binary orbits shown in
Fig.~\ref{f:binorbit} (upper panel: HD132563B; lower panel: HD132563A). 
Line styles as in Fig.~\ref{f:binorbit}.}
   \label{f:acrit}
   \end{figure}

The approach presented in Sect.~\ref{s:rvdiff} can also be used
to put constraints on the orbit of the wide components (see Hauser \& Marcy
\cite{hauser}).
We use the values of $x$, $y$, $v_x$ and $v_y$  given above and  
explore the orbital parameters resulting from the allowed values of
$z$, the separation between the components along the line of sight.
Because a range of values for the RV difference and the total mass
HD 132563A are possible, we perform the calculation for three
representative pairs of these quantities (taking the
correlation into account that results from our MonteCarlo simulations, see
Fig.~\ref{f:simul}): $|\Delta RV|$=1.2, 1.4, and 1.6 km/s and
mass of HD132563Ab of 0.20, 0.45, and 0.70 $M_{\odot}$, respectively.
The results are shown in Fig.~\ref{f:binorbit}, while Fig.~\ref{f:acrit} shows
the corresponding critical semimajor axis for dynamical stability ($a_{crit}$;
Holman \& Weigert \cite{holman})\footnote{The equation to derive $a_{crit}$ is defined
only for eccentricities below 0.8, and therefore the results shown in  Fig.~\ref{f:acrit}
are not accurate for the orbits with extreme eccentricities.}.
The plots of the orbital parameters for the three adopted configurations 
look qualitatively similar. 
Only highly eccentric orbits are allowed for $z<0$, with rather close
separations at periastron and limited regions for dynamical
stability around both components. Such orbits of the wide binary are then
excluded by the presence of  the companion orbiting HD132563A, which achieves a
separation at apoastron of at least 15 AU.
The most plausible zone is for $z$ between 100 and 700 AU, corresponding
to orbits with $a\sim 400-3000$~AU, $e\sim 0.2-0.8$, 
periastron larger than $\sim 100$~AU, and $a_{crit}$ larger than 30 AU
for both components of the wide pair.

\section{Discussion}
\label{s:discussion}

\subsection{The planet around HD132563B}
\label{s:planet}

The planet around HD132563B is a rather typical giant planet, lying in
a well populated region in planetary mass vs orbital period or semimajor axis 
diagram. The eccentricity is moderate and also not unusual for planets at
this separation.
HD 132563B has a lower metallicity with respect to the bulk of planet
hosts, but its metal deficiency cannot be considered really peculiar.

The orbit of the HD132563 system is still undetermined, so we did not perform
a detailed derivation of the dynamical stability of the planet.
However, the presence of the HD132563A pair, with a separation several times larger
than the planet from its host star, guarantees that the planet orbit is stable.

\subsection{Abundance difference between binary components with and without planets}
\label{s:diffabu}

As shown in Sect.~\ref{s:star}, there are no significant abundance differences between the
components of HD 132563 ($\Delta$ [Fe/H]$=+0.012\pm0.013$).
Our systematic searches for abundance differences in binary systems with similar components
(Desidera et al.~\cite{chem2}; \cite{chem3}) allowed us to infer that large
abundance differences are rare. Only one case with an abundance difference larger than 
0.10 dex was found (HIP 64030) but the origin of the abundance variations might be unrelated 
to accretion of planetary material because the primary of this system is a blue straggler (Desidera
et.~\cite{bs}).
Also, abundance analysis of stars in open clusters did not reveal robust evidence of
members with enriched abundances (e.g. Paulson et al.~\cite{paulson03}; Shen et al.~\cite{shen05}).
As we have only one planet known  in the sample of Desidera et al.~\cite{chem2} (HD 132563, presented
here) and the samples of Desidera et al.~\cite{chem3} and open cluster members were not systematically searched 
for planets, it is
possible that the lack of abundance anomalies is due to the paucity of planets in our sample.
We then consider in Table~\ref{t:diffabu}
the abundance difference in binaries with similar components for which giant planets
have been detected.
While optimized differential abundance analysis was only performed for 16 Cyg  and HD132563,
there are no indications of large ($\ge 0.10$ dex) abundance differences in any of the seven systems
for which data are available. These results indicate that large metallicity enhancements 
comparable to the
typical metallicity difference between stars with and without giant planets (e.g. Gonzalez \cite{gonzalez06}) are rare. 
We then provide further support to the idea  that high metallicity favors the formation
of giant planets around solar-type stars (e.g. Santos et al.~\cite{santos04}; Fischer \& Valenti \cite{fv05};
Mordasini et al.~\cite{mordasini09}).

\begin{table*}[h]
   \caption[]{Abundance differences for binary systems with similar components that host planetary companions.}
     \label{t:diffabu}
 \begin{center}
       \begin{tabular}{lcccccl}
         \hline
         \noalign{\smallskip}
System  & Planet & $T_{eff} (A)$ & $\Delta T_{\mathrm eff}$(A-B) & [Fe/H](A)  &   $\Delta$[Fe/H](A-B) & Ref. \\
        & host   &    K          &           K                    &           &                       &    \\
         \noalign{\smallskip}
         \hline
         \noalign{\smallskip}
\hline
\object{16 Cyg}                  &   B  & 5745 & $62\pm14$ & +0.10 & $-0.025\pm0.009$ & 1 \\
\object{16 Cyg}                  &   B  & 5765 &  39       & +0.05 & $ 0.00\pm0.01$   & 2 \\
\object{HD 80606}/7              &   A  & 5700 &   0        & +0.46  & $-0.01\pm0.11$   & 3 \\
\object{HD 80606}/7              &   A  & 5419 &  35       & +0.254 & $+0.002\pm0.081$ & 4 \\
\object{HD 99491}/2              &   B  & 5502 & 547       & +0.34  & $-0.02\pm0.03$   & 5$^{\mathrm{a}}$ \\
\object{HD 99491}/2              &   B  & 5650 & 400       & +0.40  & $+0.04\pm0.13$   & 3 \\
\object{HD 99491}/2              &   B  & 5557 & 654       & +0.365 & $+0.076\pm0.059$ & 4 \\
\object{HD 20781}/2              &   A  & 5774 & 518       & -0.06  & $+0.05\pm0.03$   & 6 \\
\object{HAT-P-1}                 &   B  & 6047 &  72 & +0.12  & $-0.01\pm0.05$   & 7$^{\mathrm{a}}$ \\
\object{XO-2}                   &   B  & 5500 & 160 & +0.47  & $+0.02\pm0.03$   & 8$^{\mathrm{a}}$ \\
\object{HD 132563}               &   B  & 6168 & $184\pm12$ & -0.19 & $+0.012\pm0.013$ & 9 \\

         \noalign{\smallskip}
         \hline
      \end{tabular}
 \end{center}

References: 1: Laws \& Gonzalez (\cite{laws});
            2: Takeda et al.~(\cite{takeda});
            3: Heiter \& Luck (\cite{heiter}); 
            4: Taylor (\cite{taylor05});
            5: Valenti \& Fischer (\cite{vf05}); 
            6: Sousa et al.~(\cite{sousa08});
            7: Bakos et al.~(\cite{bakos}) 
            8: Burke et al.~(\cite{burke07})
            9: Desidera et al.~(\cite{chem2})

$^{\mathrm{a}}$: Valenti \& Fischer (\cite{vf05}) actually considered the abundance differences 
                 between components of binaries with different temperatures as due to systematic effects
                 in the analysis and introduced a correction to remove them.
                 As these corrections are based on the analysis of several pairs, their impact
                 on an individual pair are likely minor.
                 This concern likely applies also to Bakos et al.~(\cite{bakos}) and Burke et al.~(\cite{burke07})
                 abundance analysis, which were performed as in Valenti \& Fischer (\cite{vf05}).

\end{table*}

Some of the pairs in Table~\ref{t:diffabu} have rather cool components for which the 
engulfment of fairly large amounts of 
rocky material are required to produce detectable abundance changes. 
The case of HD132563 is very favorable thanks to the small errors on abundance difference
and the rather thin convective zone of its components.
Following the approach by Murray et al.~(\cite{murray01}), the masses of the convective zones
of HD132563 A and B are 0.016 and 0.018~$M_{\odot}$, respectively.
The 1-sigma error on our abundance difference corresponds to the accretion of just $0.13$ of iron  for A  and $0.14~M_{\oplus}$
 for B. These quantities are smaller than the amount of rocky material that
should have been accreted by the Sun during its main sequence lifetime ($0.4~M_{\oplus}$ of iron corresponding to
 $5~M_{\oplus}$ of meteoritic material), according to Murray et al.~(\cite{murray01}).

Recent results suggest that the Sun has a peculiar abundance pattern with respect to solar
analogs, with a depletion of refractory elements relative to volatile elements 
(Melendez et al.~\cite{melendez09}; Gonzalez et al.~\cite{gonzalez10}; Ramirez et al.~\cite{ramirez10}),
possibly linked to the formation of terrestrial planets.
A more detailed evaluation of the full relative abundance pattern of HD132563 components, will be performed in a forthcoming paper,
by studying elememts
with different condensation temperature and exploring these issues.

\subsection{Planets in triple systems}
\label{s:triples}

The planet orbiting HD132563B is one of the few known in triple systems.
Table \ref{t:triples} lists the current census of planets in triple systems.
In all these systems the planet orbits the isolated component of the hierarchical triple, 
with a close stellar pair at a much larger separation. At first order, the dynamical effects of the distant pair
can be approximated by  a single star with the sum of the masses of individual components.
After deriving the critical semimajor axis for dynamical stability (Holman \& Wiegert
\cite{holman}) in this way, there are no systems for which the outer pair is found to have a strong
impact on the planetary region.
Another similar system not included in Table \ref{t:triples} is HD41004, for which
the companion of the no-planet host is in the substellar regime ($m \sin i=18~M_{J}$; Zucker et al.~\cite{zucker04}).

A different system architecture might be represented by \object{WASP-8}=CCDM 23596-3502. 
This solar type star has an M dwarf companion (\object{CCDM 23596-3502B}; mass about $0.5~M_{\odot}$) 
at 400 AU projected separation (Queloz et al.~\cite{queloz10}). Besides the transiting planet on a strongly
inclined and eccentric orbit, WASP-8 shows a long term drift of 58 m/s/yr, without any significant 
indication of curvature over two years of observations. Continuation of the observation will reveal
whether such a companion is a massive planet, a brown dwarf, or another stellar object in the system.

\begin{table*}
   \caption[]{Hierarchical triple systems with planets. The first two columns list the name and the mass of the planet host. The 3rd and 4th columns list
              the individual masses of the two components of the close pair and their semimajor axis or projected separation. The 5th column lists
              the projected separation between the planet host and the distant pair and the 6th columns the critical semimajor axis for dynamical 
              stability of planetary companions around the planet host; The 7th, 8th and 9th columns list the main planet parameters and finally
              the 10th column report the references for the individual objects.}
     \label{t:triples}
       \begin{center}

 \begin{tabular}{lccccccccl}
\hline
\noalign{\smallskip}
\multicolumn{2}{c}{Planet ~~ Host} &  \multicolumn{2}{c}{Close ~~Pair} & \multicolumn{2}{c}{Wide ~~Pair} & \multicolumn{3}{c}{Planet} & \\
\hline
Name      &  Mass &  Comp. Mass &  $\rho_{close}$ & $\rho_{wide}$ & $a_{crit}$$^{\mathrm{a}}$& M$_{\mathrm pl}$ & a$_{\mathrm pl}$& e$_{\mathrm pl}$ & Ref \\     
             &  $M_{\odot}$ & $M_{\odot}$ & AU & AU & AU & $M_{J}$  & AU & & \\
         \noalign{\smallskip}
         \hline
         \noalign{\smallskip}
30 Ari B     & 1.16  & 1.31+0.13$^{\mathrm{b}}$ & 0.02 & 1500 &  318 & 9.88 & 1.00 & 0.29 & 1,2 \\
HD 40979     & 1.19  & 0.83+0.38                & 129  & 6400 & 1452 & 3.83 & 0.85 & 0.27 & 3 \\
HD 65216     & 0.92  & 0.09+0.08                &   6  &  253 &   82 & 1.21 & 1.47 & 0.41 & 4,5 \\
HD 132563B   & 1.01  & 1.08+0.56$^{\mathrm{c}}$ &  10  &  400 &   78 & 1.49 & 2.62 & 0.22 & 6 \\
HD 178911B   & 1.06  & 1.10+0.79                &   3  &  640 &  120 & 7.35 & 0.35 & 0.14 & 7 \\
16 Cyg B     & 0.99  & 1.02+0.17                &  70  &  850 &  183 & 1.68 & 1.68 & 0.68 & 8 \\
HD 196050    & 1.15  & 0.29+0.19                &  20  &  511 &  146 & 2.90 & 2.45 & 0.23 & 5 \\
91 Aqr$^{\mathrm{d}}$    & 1.23  & 0.87+0.84    &  18  & 2250 &  461 & 2.90 & 0.30 &  --  & 9,10,6 \\
      \noalign{\smallskip}
         \hline
      \end{tabular}
 \end{center}

References: 1: Guenther et al.~(\cite{guenther09});
            2: Morbey \& Brosterhus (\cite{morbey74});
            3: Mugrauer et al.~(\cite{mugrauer07a}),
            4: Mugrauer et al.~(\cite{mugrauer07b}),
            5: Eggenberger et al.~(\cite{egg07});
            6: This paper;
            7: Tokovinin et al.~(\cite{tok00});
            8: Patience et al.~(\cite{patience02}); 
            9: Raghavan et al.~(\cite{raghavan06});
           10: WDS (Mason et al.~\cite{wds})

$^{\mathrm{a}}$: The critical semimajor axis for dynamical stability was obtained from the projected separation 
                 as in Bonavita \& Desidera (\cite{bd07}).
$^{\mathrm{b}}$: Minimum mass from spectroscopic orbit.
$^{\mathrm{c}}$: Value of plausible mass, see Sect.~\ref{s:hd132563a}.
$^{\mathrm{d}}$: Planet announced in 2003 (Mitchell et al.~\cite{mitchell03}) but never published in refereed journals, 
                 confirmed at the Conference " Planetary Systems Beyond the Main Sequence",
                 as reported in the Extrasolar Planet Encyclopedia. Stellar mass of 91 Aqr
                 obtained from param interface (Da Silva et al.~\cite{param}) using the input parameters
                 from Hekker et al.~(\cite{hekker07}); stellar masses
                 of the companions from magnitudes in WDS and mass-luminosity relation. 
\end{table*}

For five of the eight hierarchical systems in Table \ref{t:triples}, the mass of the planet host
is lower by more than 10\% than the sum of the masses of the wide stellar companions.
These are rather special cases in the current census of exoplanets in binaries, for which in nearly
all cases the planet host is the most massive star in the system.
This is likely from a combination of selection effects. In most cases only the stellar primary
is under RV monitoring \footnote{Among 210 members of multiple systems identified by 
Bonavita \& Desidera (\cite{bd07}) and Bonavita et al.~(\cite{bd10})
in the ``Uniform Detectability (UD)'' sample of Fischer \& Valenti (\cite{fv05}), 
only in 36 cases (18 pairs) were both components of the binary system observed.},
and a significant fraction of the secondaries are low-mass stars for which there is evidence of a lower frequency
of giant planets (Endl et al.~\cite{endl06}, Johnson et al.~\cite{johnson07}).
Solar-type companions of early type stars would allow investigating the role of the mass ratio
on the occurrence of planets.
Hierarchical triples represent an interesting alternative to such a program, providing
a sample of stars that are affected by the dynamical influence of the equivalent of a more
massive companion.

While a detailed study of the planets in triple systems is beyond the scope
of this paper, it seems that hierarchical triple systems do not represent a hostile
environment for planet formation around the isolated component, regardless of the mass
ratio between the planet host and the sum of the masses of the other components.
In the sample of our survey, we have detected nine additional components that are likely stellar
(19\% of the pairs), and the first planet was detected in one of such.
In the sample of planets in binaries studied in Desidera \& Barbieri (\cite{db07}), there are
four triple systems among the 35 considered in the study (11\%). This is lower than the
fraction of triple systems with respect to binaries by about a factor of two (Tokovinin \cite{tokovinin}),
but the multiplicity
estimate of the companion of planet host is underestimated (e.g. no dedicated RV monitoring looking
for short-period companions in several cases).
In the sample of Bonavita \& Desidera (\cite{bd07}, with updates in Bonavita et al.~\cite{bd10}),
there are 25 triple systems. For 12 of these systems, the component under RV monitoring is 
a member of the close pair, with no planets detected around them. This is consistent with
the paucity of planets in close binaries (see  Bonavita \& Desidera \cite{bd07}; Eggenberger \cite{egg09});
in these cases, the dynamical effects on the target are dominated by close companion. 
The other 13 systems\footnote{14, if we include HD 97344, which has a close pair of brown dwarfs (total mass $0.05~M_{\odot}$)
at a projected separation of about 2000 AU.} are those for which the isolated component is  
under RV monitoring, with three planets detected (those orbiting HD~40979, HD~178911B, and HD~196050), leading to
a frequency of planet of 23\%.
Selection biases certainly play a role in such a large frequency: dedicated
deep imaging observations such as those that allowed the binarity of the companions of HD~40979 and 
HD~196050 to be identified
were not performed for the majority of the stars in the UD sample. If we consider only the triple systems
in the UD sample identified independently on the planet detection we find a planet frequency of 9\% 
(1/11, HD~178911B), fully compatible with what is found for planets around single stars and components
of wide binaries. This result further
supports that isolated components in triple systems are not devoid of planetary companions.

In all but one stellar triple with planets, the separation of the stellar pair is larger than the
planet semimajor axis. While selection effects certainly play a role (no spectroscopic monitoring
of the companion to detect spectroscopic binaries in several cases), a moderately wide pair in a triple system
guarantees that the present stellar orbits are not disruptive for the planetary system around
the isolate component. A wide stellar triple might also indicate  a rather unperturbed
dynamical history for the system.

\section{Conclusion}
\label{s:conclusion}

We have presented the discovery of a giant planet with a projected mass $m \sin i = 1.49~M_{J}$ 
at 2.6 AU in a moderately eccentric orbit around HD132563B. 
This is the first planet detected using SARG at TNG as part of the survey that
monitored about 100 stars that are members of wide binaries with similar components.

The star has a wide companion (HD132563A) at a projected separation
of 400 AU, which we found to be  a spectroscopic binary with a period of about
25-30 yr in a highly eccentric orbit.
Taking various observational constraints into account (unresolved photometry of HD132563A, 
line profile analysis, adaptive optics imaging, astrometric data), we concluded that most likely
the companion has a mass of about $0.5-0.6~M_{\odot}$.
Therefore, HD 132563Bb is one of eight planets known to be in stellar triple systems.
A tentative statistical analysis indicates that the isolated component in a hierarchical
triple system is not hostile to planet formation, with planet frequency as high
as for single stars and the components of wide binaries.

HD132563B is a slightly metal-poor star with a mass very close to that of Sun. 
Its companion is less than 200~K warmer. Such a small temperature difference, coupled with
the small convective envelope of the star, makes this system ideal for exploring 
accretion of metal-rich material of planetary origin. The differential abundance analysis 
(already performed in Desidera et al.~\cite{chem2}) yielded no significant abundance difference,
with an error small enough to exclude accretion of about $5~M_{\oplus}$ of meteoritic material,
which is the quantity that should have been accreted by the Sun during its main sequence lifetime.

This result, along with the lack of objects in binary systems and open clusters
showing large enhancements of iron abundance that are linked to the evolution of planetary systems,
agrees with the idea that large abundance anomalies, comparable to the
typical metallicity difference between stars with and without giant planets, are rare. 

\begin{acknowledgements}

This research has made use of the 
SIMBAD database, operated at the CDS, Strasbourg, France. and of the Washington Double Star Catalog 
maintained at the U.S. Naval Observatory. \\
We thank the TNG staff for contributing to the observations and the TNG TAC for the
generous allocation of observing time.
We thank R.~Ragazzoni and A.~Ghedina for useful discussions on AdOpt@TNG.
We thank B. Mason for providing the astrometric data collected in the Washington Double Star Catalog.\\
This work was partially funded by PRIN-INAF 2008 
``Environmental effects in the formation and evolution of extrasolar planetary
systems''.
E.C. acknowledges support from CARIPARO.
We thank the anonymous referee for helpful comments.

\end{acknowledgements}

\Online

\begin{table}
   \caption[]{Differential radial velocities of HD~132563~A}
     \label{t:rva}
       \centering
       \begin{tabular}{ccc}
         \hline
         \noalign{\smallskip}
         JD -2450000  &  RV & error  \\
         \noalign{\smallskip}
         \hline
         \noalign{\smallskip}
 2013.6585    &    27.7    &     8.3   \\
 2115.4753    &    12.3    &     8.3   \\
 2327.6477    &    18.0    &    12.1   \\
 2394.6013    &    52.2    &     6.8   \\
 2446.5623    &    32.5    &    10.8   \\
 2774.5525    &    71.2    &     7.1   \\
 3129.6248    &    98.3    &     7.1   \\
 3167.5352    &    93.8    &     6.4   \\
 3574.4756    &   119.9    &     8.7   \\
 3871.6550    &   125.5    &     6.1   \\
 3898.5486    &   109.5    &     5.8   \\
 3901.5701    &   116.5    &     9.3   \\
 3960.4365    &    99.9    &     6.3   \\
 3961.4538    &   102.5    &     6.1   \\
 4099.7730    &   113.6    &     8.5   \\
 4100.7491    &   122.7    &     7.0   \\
 4160.6594    &   120.7    &     7.9   \\
 4161.6741    &   101.3    &     5.5   \\
 4190.6703    &   117.3    &     8.0   \\
 4220.6160    &   121.4    &     6.1   \\
 4221.5886    &   129.9    &     6.8   \\
 4250.5588    &   101.3    &     8.8   \\
 4251.5495    &    98.6    &     8.5   \\
 4252.5885    &    87.0    &     9.3   \\
 4276.5013    &   112.8    &     6.8   \\
 4309.4613    &    94.8    &     8.5   \\
 4311.4002    &    68.0    &     8.8   \\
 4338.3612    &    75.4    &     6.6   \\
 4339.3567    &    68.1    &     6.9   \\
 4369.3521    &   112.0    &     7.0   \\
 4515.6647    &    90.1    &     6.7   \\
 4516.6545    &    98.9    &    11.2   \\
 4544.6446    &    74.9    &     9.1   \\
 4578.5704    &    37.4    &    13.1   \\
 4579.5825    &    87.9    &     8.1   \\
 4605.5027    &    66.6    &     8.5   \\
 4606.5871    &    79.7    &    13.0   \\
 4607.5685    &    60.9    &     9.5   \\
 4609.5616    &    69.6    &     6.7   \\
 4610.5010    &    54.2    &     9.1   \\
 4636.5548    &    56.5    &     8.6   \\
 4669.5064    &    65.7    &    10.1   \\
 4670.5083    &    40.1    &     9.2   \\
 4691.4399    &    27.5    &     6.7   \\
 4849.7753    &    -4.9    &     9.3   \\
 4961.4343    &   -37.1    &     5.2   \\
 4962.4406    &   -36.8    &     7.1   \\
 5046.4187    &   -87.6    &    10.0   \\
 5054.4395    &   -87.8    &     8.9   \\
 5449.3704    &  -447.4    &    12.0   \\
 5600.7471    &  -845.1    &     7.0   \\
 5643.6802    & -1025.7    &    10.4   \\
 5658.6051    & -1061.9    &     9.4   \\
         \noalign{\smallskip}
         \hline
      \end{tabular}

\end{table}

\begin{table}
   \caption[]{Differential radial velocities of HD~132563~B}
     \label{t:rvb}
       \centering
       \begin{tabular}{ccc}
         \hline
         \noalign{\smallskip}
         JD -2450000  &  RV & error  \\
         \noalign{\smallskip}
         \hline
         \noalign{\smallskip}
 2013.6719    &    34.5    &    10.8   \\
 2115.4885    &    41.4    &    11.2   \\
 2327.6601    &    21.6    &    20.1   \\
 2394.7006    &    24.7    &     7.9   \\
 2446.5757    &     5.1    &    12.1   \\
 2774.5654    &    -8.4    &     7.0   \\
 3129.6370    &    35.1    &     8.0   \\
 3167.5469    &    19.7    &     9.6   \\
 3246.4146    &    19.9    &    10.5   \\
 3371.7370    &    26.5    &    13.5   \\
 3574.4866    &    11.2    &     9.9   \\
 3871.6669    &    27.7    &     6.8   \\
 3898.5605    &    -5.6    &     5.9   \\
 3901.5821    &   -25.9    &     9.8   \\
 3960.4494    &    -8.1    &     6.8   \\
 3961.4655    &    -8.3    &     7.6   \\
 4099.7852    &   -30.3    &    12.2   \\
 4100.7608    &   -17.8    &     9.0   \\
 4160.6721    &   -11.6    &     9.4   \\
 4161.6960    &   -39.2    &     8.1   \\
 4190.6825    &   -38.1    &     9.0   \\
 4220.6272    &   -15.3    &     8.7   \\
 4221.6003    &    -9.8    &     6.6   \\
 4250.5700    &   -28.6    &    10.7   \\
 4251.5607    &     3.8    &    11.5   \\
 4252.5992    &   -21.9    &    13.3   \\
 4276.5130    &   -30.9    &     8.4   \\
 4309.4731    &   -19.4    &    11.2   \\
 4311.4120    &   -33.1    &    10.6   \\
 4338.3729    &   -18.6    &     7.6   \\
 4339.3680    &   -20.8    &     8.8   \\
 4369.3643    &    -7.7    &     8.4   \\
 4515.6764    &    -6.2    &     8.4   \\
 4544.6558    &     2.0    &    11.3   \\
 4605.5149    &   -10.8    &    12.0   \\
 4609.5743    &    11.8    &     8.0   \\
 4669.5191    &    29.4    &    13.1   \\
 4670.5210    &    20.9    &    11.1   \\
 4691.4526    &    19.9    &     9.9   \\
 4961.4455    &    30.6    &     7.7   \\
 4962.4523    &    17.1    &     7.4   \\
 5054.4531    &    35.8    &    13.4   \\
 5434.3939    &    18.0    &    13.3   \\
 5449.3543    &    18.1    &    14.5   \\
 5600.7241    &     4.9    &     7.4   \\
 5643.7032    &   -22.2    &     9.3   \\
 5658.6217    &   -41.3    &    11.9   \\
         \noalign{\smallskip}
         \hline
      \end{tabular}

\end{table}

\end{document}